\pdfoutput=1

\documentclass[prl,superscriptaddress,amsmath,amssymb,twocolumn]{revtex4-1}
\usepackage{etoolbox}
\usepackage[utf8]{inputenc}
\usepackage{graphicx,bm,amsmath}
\usepackage[usenames,dvipsnames]{xcolor}
\usepackage[colorlinks,bookmarks=false,citecolor=NavyBlue,linkcolor=Black,urlcolor=blue]{hyperref}

\usepackage[bbgreekl]{mathbbol}
\usepackage{xcolor}
\usepackage[normalem]{ulem}
\usepackage{braket}
\usepackage{bm}
\usepackage{esint}
\usepackage{tabularx}
\def\doi{http://dx.doi.org/}
\usepackage{graphicx}
\usepackage{float}
\def\Tr{\operatorname{Tr}}
\newcommand{\be}{\begin{equation}}
\newcommand{\ee}{\end{equation}}
\newcommand{\bea}{\begin{eqnarray}}
\newcommand{\eea}{\end{eqnarray}}

\def\nn{\nonumber\\}

\usepackage{comment}

\def\barT{\hat{T}^-}

\def\nn{\mathfrak{n}}

\def\eee{{\sf e}}
 \def\noise{\eta}

\def\WW{{\sf W}}

\def\xb{\mathbf{x}}
\def\yb{\mathbf{y}}

\def\Pstat{P_{\rm stat}}

\def\hyp{{}_2 F_1}

\newcommand{\titleinfo}{Universal out-of-equilibrium dynamics of 1D critical quantum systems perturbed by noise coupled to energy}

 \AtEndEnvironment{thebibliography}{

 }

\begin{document}

\title{\titleinfo}

\author{Alexios Christopoulos}
\affiliation{Laboratoire de Physique Th\'eorique et Mod\'elisation,
CY Cergy Paris Universit\'e, \\
\hphantom{$^\dag$}~CNRS, F-95302 Cergy-Pontoise, France}    
\author{Pierre Le Doussal}
\affiliation{Laboratoire de Physique de l'\'Ecole Normale Sup\'erieure, CNRS, ENS \& PSL University, Sorbonne Universit\'e, Universit\'e de Paris, 75005 Paris, France}
\author{Denis Bernard}
\affiliation{Laboratoire de Physique de l'\'Ecole Normale Sup\'erieure, CNRS, ENS \& PSL University, Sorbonne Universit\'e, Universit\'e de Paris, 75005 Paris, France}
\author{Andrea De Luca}
\affiliation{Laboratoire de Physique Th\'eorique et Mod\'elisation,
CY Cergy Paris Universit\'e, \\
\hphantom{$^\dag$}~CNRS, F-95302 Cergy-Pontoise, France}

\begin{abstract}
We consider critical one dimensional quantum systems initially prepared in their groundstate and perturbed by a smooth noise coupled to the energy density. By using conformal field theory, we deduce a universal description of the out-of-equilibrium dynamics.
In particular, the full time-dependent distribution of any $2$--pt chiral correlation function can be obtained from solving two coupled ordinary stochastic differential equations. In contrast with the general expectation of heating, we demonstrate that the system reaches a non-trivial and universal stationary state characterized by broad distributions. 
As an example, we analyse the local energy density: while its first moment diverges exponentially fast in time, the stationary distribution, which we derive analytically, is symmetric around a negative median and exhibits a fat tail with $3/2$ decay exponent. 
We obtain a similar result for the entanglement entropy production 
associated to a given interval of size $\ell$. The corresponding stationary distribution has a $3/2$ right tail for all $\ell$, and converges to a one-sided Levy stable for large $\ell$.
Our results are benchmarked via analytical and numerical calculations for a chain of non-interacting spinless fermions with excellent agreement.
\end{abstract}

\maketitle
\emph{Introduction. --- } The coherent dynamics of macroscopic quantum systems has attracted a lot of interests in the last years, both for its fundamental importance and for its relevance to experimental setups~\cite{Gross995,Bloch2005,Schneider2012,Jepsen2020}. 
From the theoretical point of view,  recent advances have led to a much deeper level of understanding about the mechanism of equilibration and thermalisation of isolated many-body quantum systems brought out-of-equilibrium. In most cases, the eigenstate thermalisation hypothesis ensures relaxation to the canonical Gibbs ensemble and the emergence of standard thermodynamics. In recent years, the quest for intriguing out-of-equilibrium phases which escape thermalisation
has pinpointed a few phenomena, such as many-body localization (MBL)~\cite{ALET2018498}, equilibration towards generalized Gibbs ensembles due to integrability ~\cite{Calabrese_2016}, quantum scars~~\cite{turner2018weak,serbyn2021quantum,bluvstein2021controlling} and Hilbert space fragmentation~\cite{SalaPRX20,KhemaniHermeleNandkishorePRB20}. In particular, for integrable systems, anomalous transport has been observed beyond the expected ballistic, with superdiffusive behavior~\cite{PhysRevLett.123.186601, gopalakrishnan2019anomalous, PhysRevX.11.031023, bulchandani2021}. Nevertheless, weak integrability breakings have been shown to eventually lead to thermalisation~\cite{PhysRevLett.125.240604, PhysRevB.101.180302, bastianello2021hydrodynamics}.

Recently, the possibility of exploring periodically driven systems has led to a larger class of setups, which culminated in the discovery of discrete time crystalline order~~\cite{khemani2019brief,ElseReview20}. The stability of these phases is based either on MBL which prevents heating to infinite temperature 
or on high-frequency expansion which leads to long-lived prethermal behavior~\cite{MachadoPRX20}. 

Stochastic unitary dynamics in discrete time has recently also attracted a lot of interest, with several exact results involving random unitary circuits~\cite{PhysRevX.7.031016, PhysRevB.99.174205, PhysRevX.8.041019}. Although 
the finite time dynamics exhibits interesting connections to growth processes, at large time the system relaxes to a trivial infinite temperature ensemble.
Other results were obtained the context of stochastic dynamics in continuous time~\cite{PhysRevB.98.184416, SciPostPhys.3.5.033,PhysRevLett.119.110201}, on the Fredrickson-Andersen model~\cite{PhysRevB.98.195125}, and the quantum simple symmetric exclusion process (QSSEP) \cite{bernard2021can, PhysRevLett.123.080601}.
In these case, by looking at the average dynamics over noise realisations, one obtains an effective Lindbladian description, which once again admits only the infinite temperature state as stationary point. It is not obvious however that the average dynamics is representative of the typical one. 
Recent studies have thus explored the statistical behavior beyond the average dynamics~\cite{ SciPostPhys.3.5.033}, including large deviations~\cite{PhysRevLett.110.150401,10.21468/SciPostPhys.6.4.045,PhysRevLett.122.130605, Carollo2021}. In particular, for the finite QSSEP with periodic boundary conditions, it was shown that the stationary state is uniformly distributed among Gaussian states with the same occupancy of the initial state~\cite{10.21468/SciPostPhys.6.4.045}. 
In the presence of appropriate open boundary conditions, a non-trivial steady state can be attained in the QSSEP~\cite{Bernard2021,PhysRevLett.123.080601}.
Another remarkable mechanism involves the use of quantum measurements which compete with the inner unitary dynamics of the system to produce non-trivial stationary states visible in the statistics of entanglement~\cite{PhysRevX.9.031009}.
In general, an important question for quantum stochastic dynamics is whether non-trivial stationary distribution can emerge when the thermodynamic limit is taken before the large time one. 
In this direction, the finite time fluctuations were studied for QSSEP in \cite{2021arXiv210702662B}.

In this paper, we consider critical one dimensional quantum systems initially prepared in their groundstate. In practice, the starting Hamiltonian is homogeneous and gapless so that scale invariance is present. In this case, its low energy spectrum is independent on the microscopic details and is well-described by a conformal field theory (CFT). 
The behavior of quantum systems perturbed by different sources of noises has attracted great interest in the recent years~\cite{PhysRevB.86.060408, PhysRevB.89.024303, PhysRevLett.118.050402}, in particular investigating their stability properties under $1/f$ noise~\cite{DallaTorre2010, PhysRevB.85.184302}.
Here, we introduce a spatially-smooth white noise at $t>0$ coupled to the energy density and bring the system out-of-equilibrium by evolving it under the corresponding unitary dynamics.
Here, we show that the full distribution of correlation functions reaches a non-trivial stationary limit, not visible at the level of noise averages which instead exhibit apparent heating. By using conformal field theory, we deduce a universal description of the out-of-equilibrium dynamics. 
Following Ref.~\cite{Bernard_2020}, the noise coupled to the energy density is interpreted as a random metric in the CFT formulation. As a result, the dynamics of chiral operators can be solved in terms of stochastic trajectories. In~\cite{Bernard_2020} the focus was on quench protocols resulting in an initial state with short spatial correlation length, while here the initial state is gapless with quasi long-range order. By studying in detail the stochastic trajectories we show that 
any $2$--pt chiral correlation function can be obtained from solving the two coupled ordinary stochastic differential equations in \eqref{stoch1}. Remarkably, this leads to 
a stationary state characterized by broad distributions.
We analyse in particular the local energy density.
We are able to obtain the first moment at all time and show that, as a consequence of conformal invariance, it diverges exponentially fast in time; nevertheless, the average is not indicative of the typical behavior: the stationary distribution reaches a simple and universal form with a fat tail with $3/2$ decay exponent (see~\eqref{eq:Pstatsmallell}) and thus no finite integer moments.   
We obtain a similar result for the entanglement entropy production 
associated to a given interval of size $\ell$. The corresponding stationary distribution still has a $3/2$ right tail for all $\ell$, and converges to a one-sided Levy stable for large $\ell$,
see \eqref{eq:Pstatlargeell}, whose physical origin can be related to the return probability of a Brownian process. 

To test the universality of this theory, we study analytically and numerically a chain of non-interacting spinless fermions at half-filling. 
At low energy, they are well described by Dirac fermions corresponding to a $c=1$ CFT. We identify a scaling limit where the noise correlation length on the lattice diverges and the CFT predictions are recovered, as confirmed numerically by computing the local energy and entanglement entropy on the lattice.

\paragraph{Model and CFT. ---}
We consider a 1d model at a second order quantum phase transition, initially prepared in the groundstate $\ket{0}$ of its gapless Hamiltonian $\hat H_0$. To simplify the notation we will indicate the groundstate averages as $\langle \ldots \rangle = \braket{\Psi_0 | \ldots | \Psi_0}$.
To simplify, we will assume continuous space, but the treatment can be readily extended to lattice systems. At time $t= 0$, a perturbation $\hat H_1$ is turned on, by coupling a space-dependent noise term with the system energy density. The total Hamiltonian then takes the form
\begin{equation}
\label{eq:genham}
\hat{H} = \hat{H}_0 + \hat{H}_1 = \int dx (1 + \noise(x,t))\hat{h}(x)
\end{equation}
where $\hat{h}(x)$ is the hamiltonian density and $\noise(x,t)$ is the noise characterised by the space-time correlation 
$\overline{ \noise(x,t) \noise(x',t') } = \delta(t-t') f(x-x')$. 
The function $f(x)$ parameterises the noise correlation, and has the dimension of a (turnover) time; it is even and has
positive Fourier transform and we choose it smooth, monotonously decreasing for $x>0$, with a fast decay at infinity when $x \gg 1$.
In the following, we will indicate as $\bar{O}$ the average of any quantity $O$ over the noise realisations. 
An effective low-energy description of $\hat{H}_0$ can be obtained using the scale invariance which holds at the second-order critical point. 
This in turns implies an emergent conformal symmetry of the unperturbed theory~\cite{BELAVIN1984333}. A powerful description can then be obtained by means of conformal field theory (CFT), which has been successfully employed even in out-of-equilibrium dynamics and quantum quenches~\cite{Calabrese_2007, calacardy2016}.
In particular, all local operators splits into chiral ($\sim$ right moving) and antichiral ($\sim$ left moving) components organised into families~\cite{francesco2012conformal}.
All operators within each chiral family descend from a primary field $\hat{\phi}^{\pm}(x)$ with given conformal dimension $\Delta^{\pm}$. In particular, the hamiltonian density can be represented as $\hat h(x) \sim v(\hat{T}^+(x) + \barT(x))$ where $v$ is the light velocity and $\hat{T}^+$ and $\barT$ are the two components of the stress-energy tensor, with $\hat{T}^+(x) - \barT(x)$ the momentum density. This implies that under $\hat{H}_0$, chiral primary fields simply translate in time $\hat{\phi}^{\pm}(x,t) = \hat{\phi}(x \mp vt)$. In \cite{Bernard_2020}, it has been recently shown that the time evolution of primary fields under $\hat{H}$ in \eqref{eq:genham} can be interpreted as a conformal transformation. In practice, one introduces the stochastic trajectories $q^{\pm}(s)$ as solution of the Langevin equation 
\begin{equation}
\label{eq:langevin}
    \frac{dq^{\pm} (s)}{ds} = \pm v (1 + \noise(q^{\pm}(s), s) )
\end{equation}
where the Ito convention is assumed. We define the functions $X^{\pm}_t(y)$ as the initial condition for \eqref{eq:langevin} at $t = 0$ (i.e. $q^{\pm}(0) = X^{\pm}_t(y)$)
such that $q^{\pm}(t) = y$. Then, the evolution of a primary field is simply given by $\hat \phi^{\pm}(y,t) = (X_t^{\pm \prime}(y))^{\Delta^{\pm}} \hat \phi(X^{\pm}_t(y), 0)$ and arbitrary correlation functions at time $t$ can be reduced to those in the  initial state via 
\begin{equation}
\label{eq:correvol}
    \left\langle\prod_{i=1}^n \hat{\phi}_i^+(y_i, t)\right\rangle= 
        \mathcal{J}(y_1,\ldots,y_n) \left\langle\prod_{i=1}^n \hat{\phi}_{i}^+(X_t^+(y_i)) \right\rangle
\end{equation}
where the factor $\mathcal{J}(y_1,\ldots,y_n) = \prod_i (X_t^{+ \prime}(y_i))^{\Delta_i^+}$ accounts for the jacobian of the conformal transformation. An analogous equation can be written for the antichiral component with $X^+_t \to X^-_t$ and $\Delta_i^+ \to \Delta_i^-$.

To study the correlation functions in \eqref{eq:correvol} and their sample to sample fluctuations, 
we thus need the joint probability distribution function (jpdf) of the set of $2n$ random variables
$X^{\pm}_t(y_j)$, $j=1,\ldots,n$. Let us first focus on the jpdf of $X^{\pm}_t(y_j)=x_j$ for a fixed chirality, 
choosing either $\pm$, which we denote $P^{\pm}_t(\xb| \yb)$. Given $n$ trajectories $q^{\pm}_j(s)$ satisfying Eq.~\eqref{eq:langevin} 
with endpoints $q_j^\pm(0)=x_j$, $q_j^\pm(t)=y_j$, $P^{\pm}_t(\xb| \yb)$ is thus the jpdf of the initial positions 
$\xb = (x_1,\ldots, x_n)$ of these $n$ trajectories conditioned 
to the positions of their final point $\yb = (y_1,\ldots, y_n)$.
As shown in \footnote{Supplemental material. \label{supplmat}}, it satisfies the Fokker-Planck (FP) equation also studied in the context of turbulence and passive scalar~\cite{Bernard1998, 2000cond.mat..7106B, RevModPhys.73.913}
\begin{equation}
\label{eq:PFP}
\partial_t P_t^{\pm}(\xb | \yb) = ( \pm v \sum_{i=1}^n \partial_{i} + \frac{v^2}{2}\sum_{i,j=1}^n \partial_{i} \partial_{j} f(x_i - x_j)) P_t^{\pm}(\xb | \yb) 
\end{equation}
where $\partial_i = \partial/\partial x_i$. Since all trajectories are evolving according to Eq.~\eqref{eq:langevin} within the same realisation of the noise, they are correlated, which appears as an interaction in \eqref{eq:PFP}. The martingale property from the initial time implies  $\overline{x_i}  = y_i \mp v t$; additionally the trajectories cannot cross one another, so that the coordinates $\yb$ and $\xb$ are always ordered in the same way. 

\paragraph{Two-point correlations. ---}
Consider a primary field $\hat{\Phi}(x,t) = \hat{\phi}_+(x,t) \times  \hat{\phi}_-(x,t)$ with a scaling dimension $\Delta = \Delta_+ + \Delta_-$. According to \eqref{eq:correvol}, the time evolution of the 2-point correlation function satisfies
\begin{multline}
\label{eq:CfromK}
    C(y_1,y_2, t) \equiv a_0^{2 \Delta} \langle \hat{\Phi}(y_1,t) \hat{\Phi}(y_2,t)\rangle =\\= C(y_1, y_2, t = 0) e^{- \Delta_+ \kappa^+ - \Delta_- \kappa^-} 
\end{multline}
where we assume $y_1 > y_2$ and we set
\begin{equation}
\label{eq:Kvar}
    \kappa^{\pm}(y_1, y_2, t) = \ln \left| \frac{(X_t^{\pm}(y_1) - X_t^{\pm}(y_2))^2}{(y_1 - y_2)^2 X_t^{\pm \prime}(y_1)X_t^{\pm \prime}(y_2)}\right| 
\end{equation}
This expression gives access to the full statistics of the correlation function. 
We first focus on either $\kappa^+$ or $\kappa^-$.
Indeed the trajectories corresponding to the chiral and antichiral components are typically separated by a distance $\sim 2 v t$. Therefore, at large time for $2 v t \gg 1 $, 
the noises they feel become uncorrelated, and we expect the two components to decorrelate (this is discussed in more details below). 
Let us define $\ell := y_1 - y_2 > 0$ and the ratio $r = (X_t^{\pm} (y_1) - X_t^{\pm} (y_2))/\ell$. 
Using spatial homogeneity the one point pdf of $\kappa=\kappa^{\pm}(y_1,y_2,t)$ is only a function of $\ell$ and $t$
and independent on the chirality. Although this pdf does not obey a closed equation, we can derive a FP equation
for the jpdf $P_{t}(r,\kappa)$ of $\kappa$ and $r$. 

By considering four trajectories, i.e. $n=4$ in \eqref{eq:PFP}, one first obtains a FP equation for the
jpdf of $r$, $X_t^{\pm\prime}(y_1)= x'_1$ and $X_t^{\pm\prime}(y_2)= x'_2$. This is achieved through the linear change of variable 
$r=(x_1-x_2)/\ell$, $x'_1=(x_3-x_1)/\epsilon$, $x'_2=(x_4-x_1)/\epsilon$, and taking the limit $\epsilon \to 0$
(the center of mass variable $x_1+x_2$ decouples). Remarkably, performing another change of variable,
one finds that the random variables $r$ and $\kappa=\log(r^2/(x'_1 x'_2))$ are solution \footnote{Solution of \eqref{stoch1} provides the distribution for $\kappa$ in \eqref{eq:CfromK} at a fixed time $t$, but the two quantities can be different as stochastic processes in $t$. In particular, joint probabilities of $\kappa$ at different times cannot be obtained from \eqref{stoch1}.}
of the Stochastic differential equation (SDE) in Ito's form~\cite{Note1}
\be \label{stoch1} 
dr = v \, dW_1(t) \quad , \quad d\kappa = v^2 g(r) dt + v \, dW_2(t)
\ee 
where $v^2 g(r)$ is a drift term and $dW_1(t),dW_2(t)$ are two centered Gaussian white noises in time of $r$-dependent 
variances~
$\overline{dW_1(t)^2} = 2 A(r)   dt $, 
$\overline{dW_2(t)^2} = 2 C(r)   dt $ and cross correlation
$\overline{dW_1(t) dW_2(t)} = B(r)  dt$~\footnote{More precisely, the expectation value employed here is conditioned to the value of the variable $r$ at time $t$.}. Here, we introduced 
\bea \label{eq:ABC} 
&& A_\ell(r)= \frac{f(0) - f(\ell r)}{\ell^2} ~,~ B_\ell(r)=   \frac{2f'(\ell r)}{\ell}  + \frac{4(f(0)-f(\ell r))}{\ell^2 r} \;,  \nonumber \\
&& C_\ell(r)= \frac{4(f(0) - f(\ell r) + \ell r f'(\ell r))}{\ell^2 r^2} - f''(0) - f''(\ell r) \;, \nonumber \\
&& g_\ell(r)= - f''(0)-\frac{2(f(0)-f(\ell r))}{\ell^2 r^2}   \;.
\eea 
Eqs.~\eqref{stoch1} must be solved with the
initial condition $r=1$ and $\kappa=0$ at $t=0$.

Since the equation for
$r$ does not involve $\kappa$, one may first solve for $r(t)$ and 
then insert the solution for $r(t)$ in the equation for $\kappa(t)$. 
For finite $\ell$, Eq.~\eqref{stoch1} cannot be solved explicitly for an arbitrary $f(x)$. 
Nevertheless, one can understand its behavior at finite time in two regimes $\ell =  y_1 - y_2\gg 1$ and $\ell = y_1 - y_2\ll 1$,
as well as in the large time limit.

\paragraph{Small separation $\ell  \ll 1$. ---} In this case, Taylor expanding the function $f$ in \eqref{eq:ABC} one finds
the leading behavior at small $\ell$ of each function as
\bea
&& A_\ell(r) \simeq - \frac{1}{2} f''(0) r^2 \quad B_\ell(r) \simeq \frac{\ell^2}{6} f^{(4)}(0) r^3 \\
&& C_\ell(r) \simeq  -\frac{\ell^4}{72} f^{(6)}(0) r^4 \quad 
g_\ell(r) \simeq  \frac{\ell^2}{12} r^2 f^{(4)}(0)
\eea 
We see that \eqref{stoch1} can be rewritten, after a redefinition of the noises $dW_1(t)= r  dB_1(t)$ and $dW_2(t)=- \frac{\ell^2}{6} r^2  dB_2(t)$, in the form 
\be \label{eq:stoch2} 
dr = r v dB_1(t) \quad , \quad d\kappa = - \frac{\ell^2}{6} r^2 v ( - \frac{v}{2} f^{(4)}(0) dt + dB_2(t) )
\ee 
where now $dB_1(t),dB_2(t)$ are $r$-independent Gaussian white noises with fixed correlation matrix 
$\overline{dB_1(t)^2} = - f''(0) dt$,
$\overline{dB_2(t)^2} = - f^{(6)}(0) dt$,
$\overline{dB_1(t) dB_2(t)}  = - f^{(4)}(0) dt$.

Let us first discuss the marginal distribution $P_t(r)$ of $r$ in this regime $\ell \ll 1$. 
One can first solve the stochastic equation for $r$ and obtain after an application of Ito's lemma: $r(t) = e^{-\theta t + v B_1(t)}$. 
We have defined $\theta = - v^2 f''(0)/2 > 0$, which is the inverse of a characteristic time.
Alternatively, one can change variable to $\rho=\log r$, which obeys 
the stochastic equation
$d \rho = - \theta dt + v dB_1(t)$, implying that $\rho(t)$ is a Brownian motion with a drift. Hence $P_t(r)$
is a log-normal distribution for the variable $r$, with 
\begin{equation}
\label{eq:udistrib}
\overline{\ln r} = -  \theta t \;, \quad \operatorname{Var}[\ln r] =   2\theta t \;.
\end{equation}
Since $\theta>0$, this shows, interestingly, that the trajectories $X_t(y_1)$, $X_t(y_2)$ tend to get closer as time grows, a 
manifestation of the phenomenon of sticky particles observed in turbulent fluids~\cite{gawkedzki2004sticky,2014arXiv1409.6946W,le2002integration,barraquand2020large}. 
On the other hand, $\overline{r} = 1$ holds independently of time, which shows
that although the typical value of $r$ decreases to zero, the distribution of $r$ is broadening with time. Hence higher moments, such as $\overline{r^2}$, grow with time. 

Using this result it is easy to calculate the noise average of $\kappa=\kappa^\pm$ by simply averaging 
\eqref{eq:stoch2} using that $\overline{dB_2}=0$. Evaluating $\overline{r^2} \simeq e^{2 \theta t}$
and integrating over time  one finds 
\begin{equation}
\label{eq:kappaave}
    \overline{\kappa} = \frac{v^2 f^{(4)}(0) \ell^2}{24 \theta} (e^{2\theta t}-1) + O(\ell^4)
\end{equation}
This result allows to obtain the noise average of $\ln C(y_1,y_2,t)$, irrespective of
the possible correlations between $\kappa^+$ and $\kappa^-$, by averaging the logarithm
of \eqref{eq:CfromK}. It is possible to calculate the higher integer moments $\overline{\kappa^n}$ and one 
finds \cite{Note1} that they all grow exponentially with time within the validity of the small $\ell$ regime. However, upon averaging \eqref{stoch1} over the noise, we observe that $\overline{\kappa} \leq 2 \theta t$ is an exact bound 
since $g(r) \leq g(\infty) = - f''(0)$.  Thus, while the moments are still diverging at large time, 
the exponential growth is only valid when $\ell^2 e^{2 \theta t} \lesssim 1$.

Nevertheless, as we now show, the
pdf of $\kappa$ remarkably converges to a stationary distribution at large time. This distribution is
very broad and consistently does not possess any integer moments for $n \geq 1$. 
To obtain the pdf of $\kappa$, we proceed in two steps: first we plug the solution $r(t)=e^{-\theta t + v B_1(t)}$ into the equation \eqref{eq:stoch2} for $\kappa$; secondly, we use time-reversal~\cite{Note1} to recast the 
resulting stochastic equation 
in the standard form studied by Bougerol~\cite{bougerol1983exemples, vakeroudis2012bougerols}. This equation is best expressed with a change of variable from $\kappa$ to $Y$
\be 
\label{eq:kappatoY}
\kappa = \ell^2\tilde{\kappa}_0 \left( \frac{\sinh Y}{\omega_0} -1 \right) \quad , \quad \tilde{\kappa}_0= - \frac{1}{12} \frac{f^{(4)}(0)}{f''(0)}
\ee 
where we defined $\omega_0  = \bigl(\frac{f^{(6)}(0) f''(0)}{f^{(4)}(0)^2}-1
\bigr)^{-1/2}>0$. 
This leads to 
\be 
\label{eq:Ystoc}
dY = - 2 \theta \tanh Y dt + \sqrt{8 \theta} d\tilde B(t)   
\ee 
where $\tilde B(t)$ is a standard Wiener process with $d\tilde B(t)^2 = dt$. Eq.~\eqref{eq:Ystoc} describes the Langevin motion of a particle in a confining
potential $U(y) = 2 \theta \log \cosh y \simeq 2 \theta |y|$ at temperature $4 \theta$. Hence it reaches an (equilibrium) 
stationary measure at large time, $\Pstat(Y) =  C/\sqrt{\cosh(Y)}$ with $C=\sqrt{2 \pi}/\Gamma(1/4)^2$.
Thus, in the limit $\ell \to 0$, the scaled variable $\omega \equiv \sinh Y = \omega_0 ( 1 + \kappa/(\ell^2 \tilde\kappa_0))$ is described by the stationary distribution
\begin{equation}
\label{eq:Pstatsmallell}
    \mathcal{B}(\omega) \equiv 
    \frac{C}{(1 + \omega^2)^{3/4}}
\end{equation}
Consistently, the power-law tail $\propto |\omega|^{-3/2}$ implies that $\omega$ (and $\kappa$ as well) does not have finite integer moments. 

We remark that, because of the time-reversal transformation, the stochastic process \eqref{eq:Ystoc} for $Y$ 
and the original one for $\kappa$ in \eqref{eq:stoch2} are not equivalent: the former is ergodic in time, while the latter has a finite (random) limit $\kappa(t \to \infty)$. Nevertheless, they are equivalent in law at fixed $t$.

\paragraph{Large interval $\ell \gg 1$. ---} 

The leading behavior at large $\ell$ of each function up to $O(1/\ell^2)$ reads
\bea
&& A_\ell(r) \simeq \frac{f(0)}{\ell^2}  \quad B_\ell(r) \simeq 4 \frac{f(0)}{\ell^2 r}  \\
&& C_\ell(r) \simeq   - f''(0) + 4 \frac{f(0)}{\ell^2 r^2}  \quad 
g_\ell(r) \simeq - f''(0) - \frac{2 f(0)}{\ell^2 r^2} \nonumber 
\eea 
where we assume that $f(x)$ decays faster than a power law. 
At leading order one can set $B_\ell(r) 
\simeq 0$ which implies that the equation \eqref{stoch1} for $\kappa$ becomes 
independent of $r$. Using $g_\ell(r) \simeq - f''(0)$ and $C_\ell(r) \simeq   - f''(0)$
to leading order one obtains
\be \label{reslargeell} 
\kappa = 2 \theta t +  \sqrt{4 \theta } W(t)    \;, \quad r = 1 + \frac{\sqrt{2 f(0)}v}{\ell} \tilde{W}(t) 
\ee 
where $W(t), \tilde W(t)$ are two uncorrelated standard Brownian motion with $\overline{dW(t)^2}=\overline{d\tilde W(t)^2}=dt$. Note that the growth 
of $\overline{\kappa}$ saturates the exact bound $\overline{\kappa} < 2 \theta t$.
To obtain this result we have assumed not only that $\ell \gg 1$ but also that $\ell r \gg 1$.
Although this condition holds for finite time, since $r$ is undergoing diffusion, we see from \eqref{reslargeell} that for $t \sim \ell^2/f(0)$, $r(t)$ may become close to zero and the condition is violated.

\paragraph{Large-time limit. ---}

To search for a stationary distribution for any $\ell$, we derive in \cite{Note1} an evolution equation for the characteristic function of $\kappa$,
$Q_k(r_0,t) = \overline{ e^{- i k  \kappa }}^{r_0}$, where the superscript $r_0 = r(t=0)$ indicates the initial condition for the variable $r$ in \eqref{stoch1}, ultimately setting $r_0 = 1$. Looking for a time dependent solution in the large time limit $ Q_k(r_0,t) \to Q_k(r_0)$,
we obtain that 
\be 
\label{eq:QGrel}
Q_k(r) = \left(\frac{\ell^2 r^2 f''(0)}{2 (f(\ell r) - f(0))}\right)^{i k} G_k(r) \;.
\ee 
where $G_k(r)$ satisfies the Schrodinger--like equation for $r \geq 0$
\be
\label{eq:GSchro}
  - G_k''(r)  - k (k+i) V(r) G_k(r) = 0 
\ee
with boundary conditions are $G_k(0) = 1$ and $\lim_{r \to +\infty} G_k(r) = 0$ and
the potential 
\be
V(r)  
= - \frac{d^2}{dr^2} \log[f(0)-f(\ell r)] + \frac{\ell^2 f''(0)}{f(0)-f(\ell r) } \;.
\ee
Studying this equation (see \cite{Note1}), we find that the stationary distribution $\Pstat(\kappa)$ (obtained by Fourier inversion of
$Q_k(r=1)$) depends non-trivially on $\ell$ and $f(x)$, with however a symmetry valid in all cases 
\be 
\frac{\Pstat(-\kappa_0 + \beta)}{\Pstat(-\kappa_0 -\beta) } = e^{\beta} \;, \quad \kappa_0 =   - \log\Bigl( \frac{2 (f(\ell) - f(0))}{\ell^2  f''(0)} \Bigr) 
\label{eq:fluctuation}
\ee 
reminiscent of a fluctuation theorem~\cite{PhysRevLett.74.2694}. 
Beyond the asymptotic solution at small $\ell$ given in \eqref{eq:Pstatsmallell}, one can also derive at large $\ell \to +\infty$, $\kappa = \theta \ell^2 \chi / f(0) + O(\ell)$ with $\chi$ distributed according to
\begin{equation}
\label{eq:Pstatlargeell}
\mathcal{L}(\chi) \equiv \frac{1}{\sqrt{2\pi}}\frac{ e^{-\frac{1}{2 \chi}}}{\chi^{3/2}} \Theta(\chi)
\ee 
i.e. to the stable one sided Levy distribution of index $1/2$.
Eqs.~\eqref{eq:Pstatsmallell} and \eqref{eq:Pstatlargeell} characterise asymptotic behavior of the stationary distribution for $\kappa$ at small and large $\ell$ respectively.
For intermediate values of $\ell$, an explicit expression is not available for generic $f(x)$. However we provide some cases which are explicitly solvable for any $\ell$, e.g. for $f(x)=1/\cosh^n x$ with $n=1,2$.

One can show that for  sufficiently smooth $f(x)$, $Q_k(r) = 1 + O(\sqrt{k})$ at small $k$, irrespectively of $\ell$. This translates onto a $-3/2$ power--law tail on the positive $\kappa$ side. For the left tail $\kappa \to -\infty$, it follows from Eq. \eqref{eq:fluctuation} that $\Pstat(\kappa)$ decays exponentially for any for $\ell>0$.

An intuition on why the $3/2$ exponent appears is as follows: the random variable $r$ is attracted to $r=0$ since for $r<\ell$, $\log r$ is approximately Brownian with a negative drift~(see \eqref{eq:udistrib}). In this regime, $\kappa$ remains almost constant as $d\kappa \propto r^2 dt$. However, $r$ starts from $1$ and has a finite probability to move right towards $r>\ell$; in this case, $\kappa$ 
increases by $\sim \Delta t$ (see Eq.~\eqref{reslargeell}) where $\Delta t$ is the time interval before $r(t)$ hits again $\ell$. Since in this regime, $r(t)$ is approximately an unbiased Brownian, $\Delta t$ is the corresponding first-passage time, distributed as $1/(\Delta t)^{3/2}$.

Finally, let us recall for time $t \gg 1/(2v)$ the two chiral components $\kappa^+$ and $\kappa^-$ decorrelate
hence their joint distribution reaches in the large time limit the factorized form $\Pstat(\kappa^+)
\Pstat(\kappa^-)$.  We expect that $\kappa^{\pm}(t)$ defined in Eq.~\eqref{eq:Kvar} as a stochastic process in time has $\Pstat(\kappa^{\pm})$ as the one-time ergodic measure. This is not in contradiction with the fact that $\kappa$ defined in the auxiliary stochastic system \eqref{stoch1} has a (random) finite limit $\kappa^{\pm}(t \to \infty)$. 

\paragraph{Entanglement production. ---}
An interesting application of the above results is the calculation of the entanglement entropies. Let us define as $\rho_{A,t}$ the reduced density matrix for the interval $A=[y_1,y_2]$ at time $t$. Then, the Renyi entropies are defined as $\mathcal{S}_t^{(n)} = 1/(1-n)\ln \Tr \rho_{A,t}^n$. Introducing the twist fields $\hat{\Phi}_n(y,t)$~\cite{calabrese2009entanglement}, one can identify $\Tr \rho_{A,t}^n \propto \langle \hat{\Phi}_n(y_1, t) \hat{\Phi}_n(y_2, t)\rangle$. We can thus express the \textit{entropy production} $S_t^{(n)} \equiv \mathcal{S}_t^{(n)} - \mathcal{S}_{t=0}^{(n)}$ using \eqref{eq:CfromK} as
\begin{equation}
\label{eq:renentropy}
S^{(n)}_t = \frac{(n+1) c}{24 n} (\kappa^+ + \kappa^-)
\end{equation}
where we used that for twist fields $\Delta^{\pm} =  c (n - 1/n)/24$.
The Von Neumann entropy corresponds to $n=1$ and we denote it simply as $S_t$. 
Eq.~\eqref{eq:renentropy} shows that all the Renyi entropies are controlled by the same random variable, and that the details of the model enter only in the prefactor via the central charge. From \cite{PhysRevA.78.032329}, it implies that the entanglement spectrum retain the same form in each noise realization,
i.e. the density of $\log \lambda/\log \lambda_{\rm max}$ is independent of the noise, while $\log \lambda_{\rm max}=- c/24(\kappa^+ + \kappa^-)$. 
Using that the noise average $\overline{S^{(n)}_t} = \frac{(n+1) c}{12 n} \overline{\kappa}$, we see that at early times two different growth regimes exists: for large intervals ($\ell \gg 1$)
the average entropy production grows linearly with time as in \eqref{reslargeell}, while for small intervals it grows as 
in \eqref{eq:kappaave}. Finally, at large time, we find that the entropy production \eqref{eq:renentropy} reaches a stationary distribution given up to a scale by
the convolution $\Pstat * \Pstat$
determined above, still with a $-3/2$ power law tail and no integer moments.

\paragraph{Distribution of the energy density. ---}
An interesting quantity to look at is the dynamics of the energy density, which, in the CFT mapping, is encoded in the stress energy tensor $\hat{T}^+(x)$ and $\barT(x)$. Their time evolution can be obtained either by direct calculation of the commutators $[\hat{H}, \hat{T}^{\pm}(x)]$ in the Heisenberg equation \cite{Note1}.
Alternatively, one can use the fact that time evolution can be seen as a conformal mapping and the corresponding transformation of the stress energy  reads (for simplicity, we omit the $\pm$ superscript)
\begin{equation}
\label{eq:Ttransf}
\hat{T}(y,t)  = |X^{\prime}_t(y)|^2 \hat{T}(X_t (y), t =0) - \frac{c}{24\pi} (\mathcal{S}\cdot X_t)(y)
\end{equation}
where the second term 
is proportional to the central change $c$ and the Schwarzian derivative
\begin{equation}
 (\mathcal{S}\cdot X_t)(y) = \frac{2 X'''(y) X'(y)-3 X''(y)^2}{2 X'(y)^2} 
\end{equation}
Since in the initial state $\langle \hat{T}(y,0) \rangle=0$, one has simply
\be
\label{eq:TSchwarz}
\langle \hat{T}(y,t) \rangle = - \frac{c}{24 \pi} ({\cal S}.X_t)(y) 
\ee 
The time evolution of this quantity can be simply obtained from the above analysis of $\kappa^{\pm}$ in the regime $\ell \to 0$. Indeed,
expanding \eqref{eq:Kvar} in small $y_1-y_2$ we see that
\be 
\label{eq:kappaTrel}
\lim_{\ell \to 0} \frac{1}{\ell^2} \kappa(y_2+\ell,y_2,t) = - \frac{1}{6} ({\cal S}.X_t)(y_2) \;.
\ee 
Eq.~\eqref{eq:kappaTrel} is well-known in the CFT context: it reflects the occurence of the conformal anomaly in the transport of the stress-tensor in a gaussian free field theory, see~\cite{francesco2012conformal}. Here, it implies that the distribution $T^{\pm}$ can be obtained from the one of $\kappa^{\pm}$ in the limit of small $\ell$. In particular, we observe a remarkable identification between the entanglement of an infinitesimal interval and the local energy density. Indeed, denoting the time dependent local energy density as the quantum expectation $h(x,t) \equiv \langle \hat h(x, t) \rangle = v (\langle \hat{T}^+ \rangle + \langle\hat{T}^- \rangle)$, one has $h(x,t) = \lim_{\ell\to0} v S_t / (\pi \ell^2)$. Additionally, denoting the noise average (which
is space independent) as 
${\eee}(t) \equiv \overline{h(x,t)}$, one has from \eqref{eq:kappaave}
\begin{equation}
\label{T1moment}
 {\eee}(t) = \frac{c v^3\ f^{(4)}(0)}{48 \pi \theta}\Big( e^{2 \theta t}-1 \Big) \;. 
\end{equation}
For $t \gg 1/(2 v)$ we expect $\langle T^\pm \rangle$ to decorrelate; thus from \eqref{eq:kappatoY}, we extract the exact stationary distribution of the one-point energy density 
\begin{equation}
    \label{eq:energyOmega}
    \lim_{t \to \infty} h(x,t) \stackrel{\text{\tiny in law}}{=}\frac{ v c \tilde{\kappa}_0}{4\pi} (\Omega / \omega_0 - 2)
\end{equation}
where $\Omega = \omega^+ + \omega^-$ and $\omega^{\pm}$ are independent  random variables both distributed according to $\mathcal{B}(\omega)$ in \eqref{eq:Pstatsmallell}.

\begin{figure}[t!]
\includegraphics[width=0.50\textwidth]{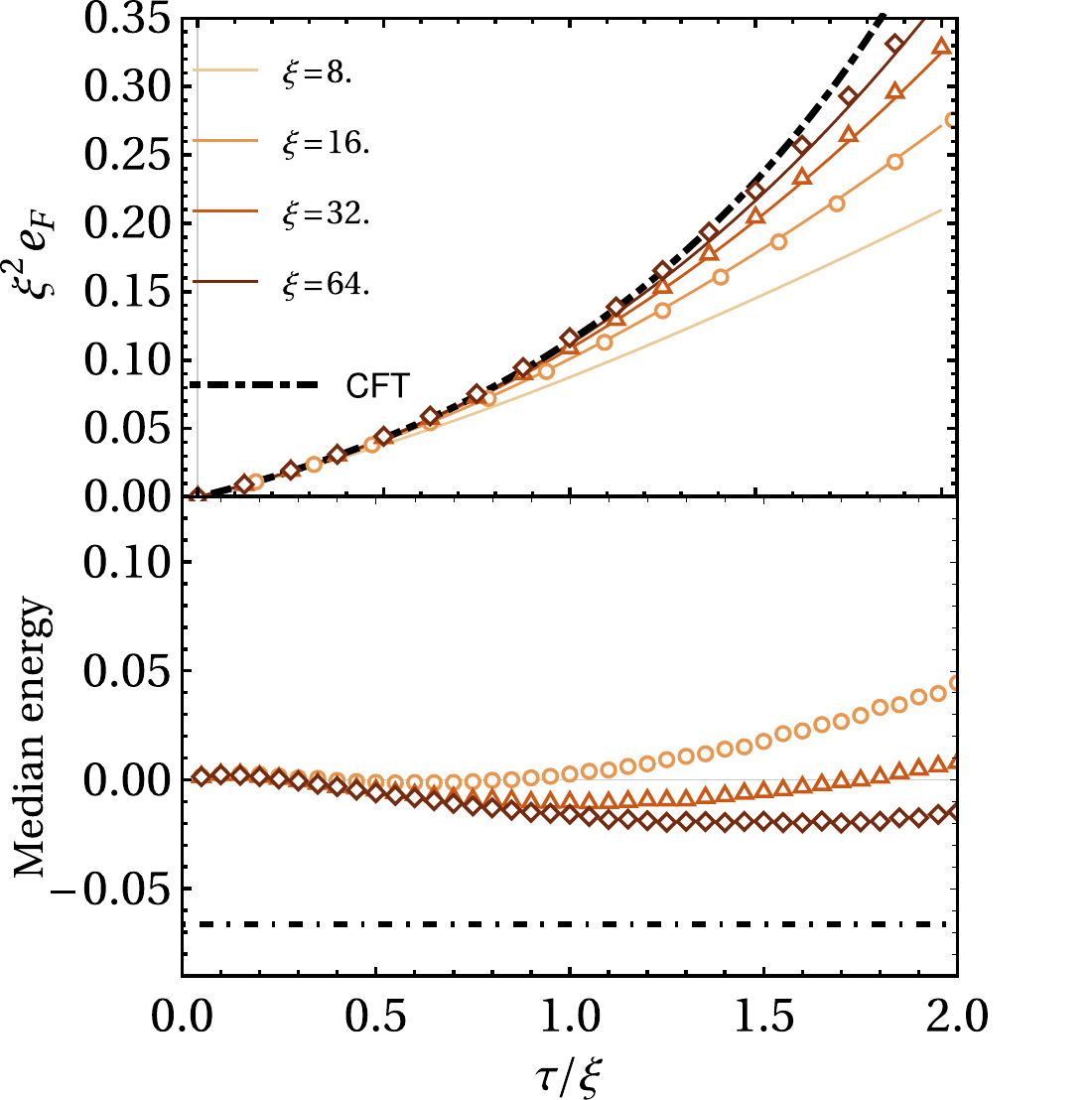}
\caption{
\label{fig:aveeneFF}
Top: The average energy $\eee_F(\tau)$ vs $t = \tau/\xi$, the scaled time, for different values of the noise correlation length $\xi$. Continuous line are obtained from the numerical solution of the Wigner function equation \eqref{eq:ndephhomo1}, while the markers corresponds to the exact dynamics of \eqref{eq:Hff} for $L = 2048$. The dotted dashed line is the CFT result \eqref{T1moment}, which from 
\eqref{eq:FFenelimit} is predicted to hold for large $\xi$. Bottom: The median of the distribution of $\xi^2 (\langle \hat h_i \rangle_{\tau} - \langle \hat h_i \rangle_0) $ vs the scaled time $\tau/\xi$. In the limit of $\xi \to \infty$, the median is expected to decrease towards the negative asymptotic value predicted by CFT (dot dashed horizontal line). For finite $\xi$, the median starts to grow at large times, suggesting that heating may eventually dominate on the lattice.
}
\end{figure}

\paragraph{Comparison with free fermions. ---}
We consider a model of non-interacting spinless fermions, where we denote 
$\tau$ the time in the lattice model with Hamiltonian
\begin{align}
\label{eq:Hff}
\hat{H}_F &= \sum_i (1 + \eta_i(\tau)) \hat{h}_i \;, \\
\hat{h}_i &= -J (\hat{a}^\dag_i \hat{a}_{i+1} + \hat{a}^\dag_{i+1} \hat{a}_{i}
) \;. \nonumber 
\end{align}
Note that the local energy density is only defined up to a total derivative. 
A good way to fix such an ambiguity and preserve the continuous limit is to impose parity
which is verified in \eqref{eq:Hff}~\cite{10.21468/SciPostPhys.6.4.049, Bernard_2016, PhysRevB.88.134301}. On the contrary, in the presence of a finite chemical potential, a different definition is required~(see \cite{Note1}).
In the absence of noise, this corresponds to a dispersion relation $\epsilon(k) = - 2 J\cos(k)$. For the noise, we choose the correlation
\begin{equation}
\label{eq:noisecorrFF}
   \overline{ \eta_i(\tau) \eta_j(\tau')} = \tau_0  \delta(\tau-\tau') F(i- j) \;, \quad F(j) = f(j/\xi)
\end{equation}
where $\xi$ is the characteristic correlation length of the discrete model and in the numerics
we choose $f(x) = 1 / \cosh(x)$, as it corresponds to an analytically solvable case. The dynamics induced by \eqref{eq:Hff} is better studied in terms of the noise averaged Wigner function $n_\tau(k) = \sum_{j'} \overline{\langle \hat{a}_{j + j'}^\dag \hat{a}_{j} \rangle}_\tau e^{i k j'} $ where $\langle \ldots \rangle_\tau$ denotes the quantum average at time $\tau$ under the evolution $\hat{H}_F$ in \eqref{eq:Hff}. 
We choose the initial state as the groundstate so that
$n_\tau(k)$ does not depend on the lattice site $j$. Also,
$n_{\tau=0}(k) = \Theta(k + k_f)  - \Theta(k - k_f)$, where $\Theta(z)$ is the Heaviside function and $k_f$ is such that $\epsilon(k_f) = 0$ and $k_f = \pi/2$, which corresponds to half filling. The system is critical and can be described with a conformal field theory with central charge $c = 1$~\cite{giamarchi2003quantum}. 
Using the Wigner function, we can express the noise averaged energy density with respect to the groundstate
 \begin{equation}
 \eee_F(\tau)  \equiv \overline{\langle \hat{h}_i \rangle_\tau-\langle \hat{h}_i \rangle_0}  =  \int \frac{dk}{2\pi} \epsilon(k) (n_\tau(k)-n_0(k))
 \end{equation}

One can derive~\cite{Note1} an exact evolution equation for $n_\tau(k)$ which reads
\begin{equation}
\label{eq:ndephhomo1}
 \partial_\tau n_\tau(k)= 
   \tau_0 \int  \frac{dk'}{2\pi}
 \tilde{F}(k') 
    \epsilon(k + k'/2)^2 (n_\tau(k + k') -  n_\tau(k)) 
 \end{equation}
The study of this equation \cite{Note1} 
shows that around the Fermi points $n_\tau(k)$ takes the
scaling form $n_\tau(k)= \nn((k_f\mp k) \xi,t=\tau/\xi)$.
In turns this leads to the result for fixed $t$ as $\xi \to +\infty$
\be 
\label{eq:FFenelimit}
\lim_{\xi \to \infty} \xi^2 \eee_F(t \xi) = {\eee}(t) 
\ee 
where ${\eee}(t)$ is given in Eq.~\eqref{T1moment}. Hence the CFT
predicts, upon rescaling, the mean energy for the fermion system.

This explicit calculation suggests that in the scaling limit of large $\xi$, the noisy dynamics in \eqref{eq:Hff} is fully captured by the universal description provided by the CFT, upon rescaling space and time as
$j = x \xi$, $\tau = t \xi$ and set $\tau_0 =  \xi$.

\begin{figure*}[t]
\includegraphics[width=0.32\textwidth,trim={0.cm 0.cm 1.2cm 0},clip]{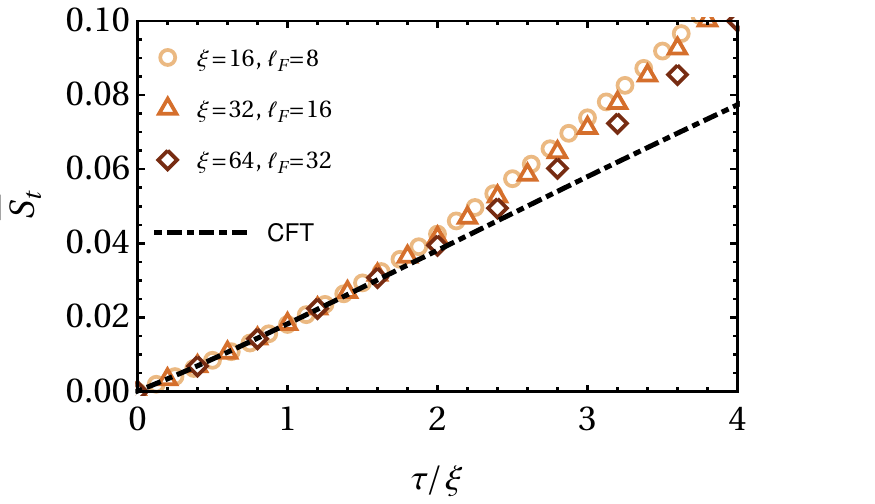}
\includegraphics[width=0.325\textwidth,trim={0.cm 0.1cm 1.cm 0.cm},clip]{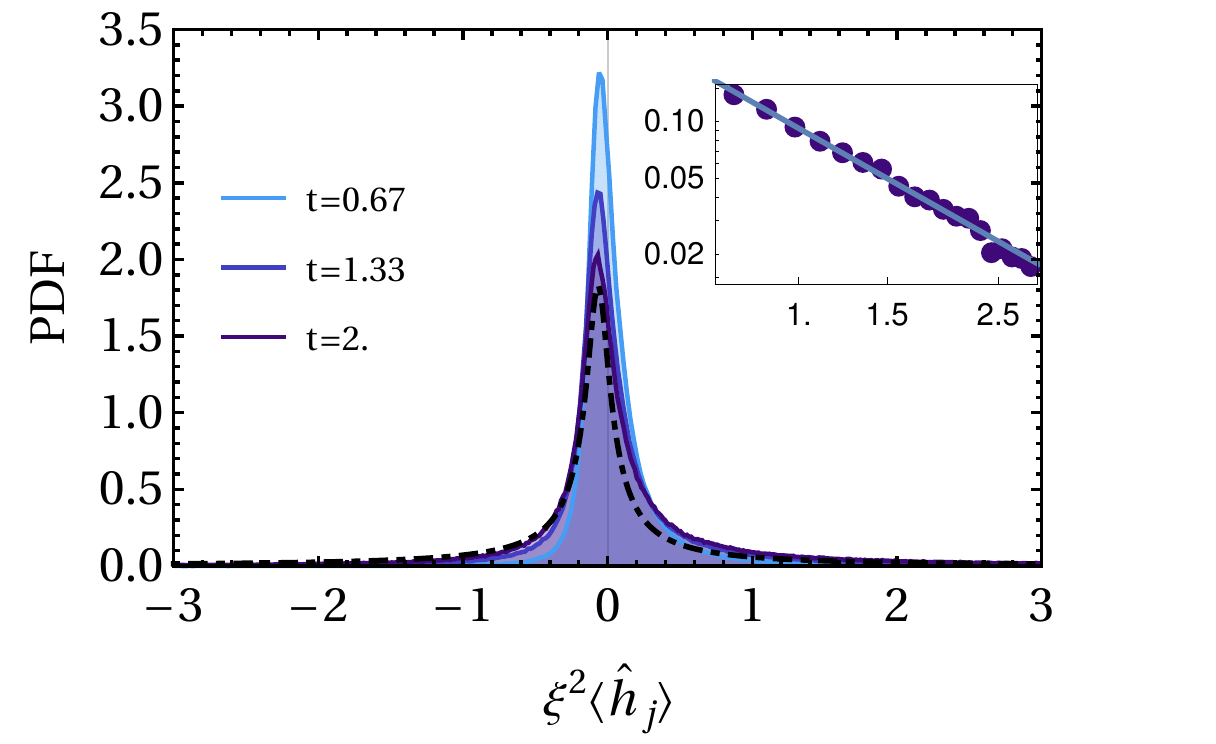}
\includegraphics[width=0.32\textwidth,trim={0.cm 0.cm 1.2cm 0},clip]{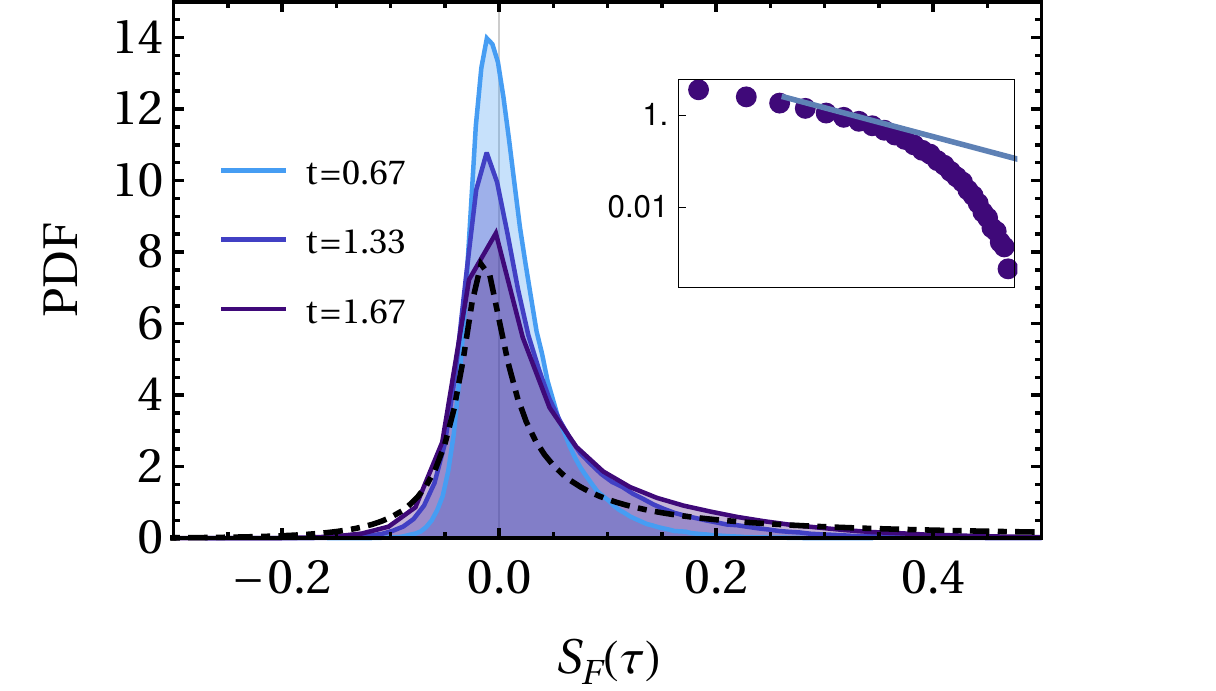}
\caption{Left: Noise--averaged Von Neumann entanglement entropy production $\overline{S^{(1)}(\tau)}$ vs $t = \tau/\xi$ evolving in time under \eqref{eq:Hff}, for increasing values of $\xi$ and fixed ratio $\ell_F/\xi = \ell = 1/2$. The dotted line is obtained from the numerical solution of Eqs. \eqref{stoch1} and using \eqref{eq:renentropy} for $n \to 1$, i.e.
$\overline{S_F(\tau)} = c \overline{\kappa}/6$ with $c=1$. Middle: Distribution of the scaled energy density $\xi^2 \langle \hat{h}_i(\tau) \rangle$ at $\xi = 64.$. For the analytical prediction, we used that in the scaling limit $\xi^2 \langle \hat{h}_j \rangle \to h(x,t)$ and at large time $h(x,t)$ is distributed as \eqref{eq:energyOmega}, which in the present case reduces to $h \stackrel{\text{\tiny in law}}{=} (3 \Omega -5)/(24 \pi)$. In the inset, the right tail of the distribution is shown in log--log scale, showing the predicted $\propto h^{-3/2}$ tail. Right:
Distribution of the entanglement entropy  $S_F(\tau)$ at $\xi = 64.$ for an interval of size $\ell_F = 32$. For the analytical prediction, we used that in the stationary limit $\lim_{\tau\to \infty} S_F(\tau) \stackrel{\text{\tiny in law}}{=} c (\kappa^+ + \kappa^-)/12$, with $\kappa^{\pm}$ independently distributed according to $\Pstat(\kappa)$. The stationary distribution $\Pstat(\kappa)$ is obtained by numerically inverting its Fourier transform $Q_k(1)$ for $\ell = 1/2$, defined in \eqref{eq:QGrel} and \eqref{eq:G2f1} of \cite{Note1}. 
All simulations are performed on systems of total length $L = 2048$ and are repeated for $N_{\rm sample} =800$ samples.
\label{fig:panel}
}
\end{figure*}

In order to validate this hypothesis, we have computed numerically the two-point correlation matrix $C_{ij}(\tau) \equiv \langle a_i^\dag  a_j  \rangle_\tau$.
Since the model \eqref{eq:Hff} is non-interacting and the initial groundstate is Gaussian, for each realisation of the noise, all quantities can be expressed via the Wick theorem in terms of the coefficients $C_{ij}$. In Fig.~\ref{fig:aveeneFF} top, we show the convergence for $\xi \to \infty$ of the noise-averaged energy density $\eee_F(\tau)$
to its CFT prediction, consistently with \eqref{eq:FFenelimit}. We observe the emergence of a characteristic time scale $\tau^\ast(\xi)$, diverging with $\xi$, after which the CFT description breaks down. A simple argument based on the broadening of the Wigner function around the Fermi points suggests $\tau^\ast(\xi) \propto \xi \ln \xi$~\cite{Note1}, which is in agreement with our numerics.

The correlation matrix $C$ can also be used to compute explicitly the Renyi entropies for any interval $I$ of size $\ell_F$. Indeed, setting $C_I$ as the $\ell_F \times \ell_F$ matrix obtained restricting the indexes of $C(\tau)$ to $I$, we have for the Von Neumann entropy of the interval $I$: $\mathcal{S}_F(\tau) \equiv - \Tr[ C_I \ln C_I + (1 - C_I) \ln (1 - C_I)]$.
In Fig.~\ref{fig:panel} left, we show the noise averaged entanglement entropy production $S_F(\tau) \equiv \mathcal{S}_F(\tau) - \mathcal{S}_F(0)$ for the fermion system for intervals of various sizes $\ell_F$ on the lattice. Our prediction is that it should equal at large $\xi$ the CFT value $\overline{S^{(1)}_t}$ (without any prefactor) with $\ell = \ell_F/\xi$. Once again, a good agreement is found and corrections emerge at $\tau \gtrsim \tau^\ast(\xi)$.

Our prediction is that the CFT describes also the distribution of these quantities. We show in Fig. \ref{fig:panel}  middle, the one-point PDF for the local energy density 
$\xi^2 \langle \hat h_i \rangle_\tau$ at the largest time $\tau$ available. As one sees it compares
reasonably well, with no free parameters, with the prediction from the CFT, i.e. the convolution $P_{\rm stat} * P_{\rm stat}$ 
where $P_{\rm stat}$ was obtained in \eqref{eq:Pstatsmallell}. 
This confirms that with the chosen $f(x)$, at this observation time, $2 v_F \tau/\xi$ is large enough, so that the two chiral components have decoupled.
As we see on Fig.~\ref{fig:aveeneFF} top, the average energy grows with time,
consistent with $P_{\rm stat}$ having infinite first moment. The median of the energy distribution, shown in Fig.~\ref{fig:aveeneFF} bottom, is thus a better probe of the typical behavior. 
Remarkably, it is found to {\it decrease} with time, approaching at large $\xi$  a stationary value compatible with the CFT prediction
$e^{\rm median}_{ \rm stat}= - \frac{c}{2 \pi} \tilde \kappa_0 < 0$, see Eq.~\eqref{eq:kappatoY}. However, for any finite $\xi$, we expect lattice effects to eventually break the CFT description. Although we have not studied it in detail, it is expected that the ultraviolet cutoff induces {\it heating} and stationarity does not hold anymore, as hinted by the rebounce in the median observed at larger times. 
Finally, in Fig.~\ref{fig:panel} right, as a representative of the finite-$\ell$ behavior, we compare the distribution of the entanglement entropy at different times for intervals of size $\ell_F = \ell \xi$, and $\ell = 1/2$, with the analytic prediction obtained from CFT.

\paragraph{Conclusion. ---}
We have identified an out-of-equilibrium protocol which leads to a non-trivial stationary state for generic gapless one-dimensional system. 

Several questions and directions remain open. First, it would be of great interest to obtain the full space-time statistics of the local energy, or of any other local operator, beyond the one-point distribution,
especially since the latter exhibit heavy tails. It also remains a challenge to extend the present method to four-point (and higher) quantum correlation functions, thus providing a full characterization of the quantum state. 
Although we focused on the infinite system, coupling between the chiral components becomes relevant at finite volume and can modify the behavior of the system at large times. 

From a more concrete perspective, it would be interesting to test the theory at other quantum critical points beyond non-interacting systems, in particular to observe the role played by interactions. 

Finally, the surprising existence of a stationary state in our setup, raises the question about the fundamental ingredients to observe similar phenomenology in other quantum stochastic systems. It is possible that the stationary state that we identified within the continuous field theory description corresponds to a long-lived prethermal state that delays heating in the corresponding lattice system, similarly to what has been observed in the context of many-body quantum scars~\cite{PhysRevX.10.021041}. More generally, it remains an open question whether lattice effects or the presence of finite correlation length are compatible with the emergence of non-trivial steady states in the thermodynamic limit.

\paragraph{Acknowledgments. ---}
PLD acknowledges support from the ANR grant ANR-17-CE30-0027-01 RaMaTraF. DB acknowledges support from the project `ESQuisses', ANR-20-CE47-0014-01.

\bibliography{biblio} 

 \onecolumngrid
\newpage 

\appendix
\setcounter{figure}{0}
\renewcommand{\thetable}{S\arabic{table}}
\renewcommand{\theequation}{S\thesection.\arabic{equation}}
\renewcommand{\thefigure}{S\arabic{figure}}
\renewcommand{\thesection}{\arabic{section}}
\renewcommand{\thesubsection}{\thesection.\arabic{subsection}}
\renewcommand{\thesubsubsection}{\thesubsection.\arabic{subsubsection}}
\setcounter{secnumdepth}{2}

\begin{center}
{\Large Supplementary Material \\ 
\titleinfo
}
\end{center}
In this supplementary material we provide additional details about the calculations in the Letter.

\tableofcontents

\section{Fokker-Planck equation}

In this section we derive the Fokker-Planck equation \eqref{eq:PFP} of the text, for the joint PDF of the 
backward stochastic trajectories $x_1=X^+_t(y_1),\dots,x_n=X^+_t(y_n)$ associated to the Langevin equation \eqref{eq:langevin}. 
We thus consider only a given chirality, here we choose $+$, but the same Fokker-Planck equation holds for the chirality $-$, with $v \to - v$. 
We do not consider here the joint PDF of both chiralities. So here we denote simply $X_t^{+} \to X_t$.

One can show that Eq.~\eqref{eq:langevin} translates into a stochastic equation for the variable $X_t(y_i)$ as a function of $t$ which takes the form (see Eq. (58-59) in Supp. Mat. of \cite{Bernard_2020})
\begin{equation}
\label{eq:dXeq}
dX_t(y_i) = \frac{v^2}{2}X''_t(y_i) f(0) dt - X'_t(y_i) v \big(dt+d\WW_t(y_i)\big)\;
\end{equation}
where $dX_t(y_i)= X_{t+dt}(y_i)-X_t(y_i)$. The $\WW_t(y_i)$ are mutually correlated Wiener processes in time $t$, which relates to the noise in Eq.~\eqref{eq:genham} and \eqref{eq:langevin} via 
\begin{equation}
\label{eq:dWdef}
\WW_t(y) = \int_0^t ds \, \eta(y, s) \;, \quad \overline{d\WW_t(y) d\WW_t(y')} = dt f(y - y')
\end{equation}
Here $d\WW_t(y)= \WW_{t+dt}(y)-\WW_t(y)$ and 
$f(y)$ is the noise correlation function defined in the text. 

Consider an arbitrary smooth function of $n$-variables $G(x_1,\ldots, x_n)$. In the case of these variables being the backward stochastic trajectories $x_i = X_t(y_i)$ of \eqref{eq:langevin}, we define:
\begin{equation}
    g_t(y_1  \dots y_n)= G\big(X_t(y_1) \dots X_t(y_n) \big)
\end{equation}
By Ito calculus the time variation $dg_t=g_{t+dt}-g_t$ of this observable is obtained by expanding up to second order.
\begin{equation}
dg_t = \sum_{j=1}^n \partial_j G\big(X_t(y_1), \ldots, X_t(y_n)\big) dX_t(y_j) + 
 \frac12 \sum_{j,m=1}^n \partial_j \partial_m G \big(X_t(y_1), \ldots, X_t(y_n)\big) dX_t(y_j)dX_t(y_{m})
\end{equation}
Here we shortened the notation setting $\partial_{x_j} \dots  \partial_{x_m} G(x_1 \dots x_n)=\partial_j \dots  \partial_m G(x_1 \dots x_n)$. Using \eqref{eq:dXeq} and \eqref{eq:dWdef}, we derive
\begin{align}
& dX_t(y_i)\, dX_t(y_j) =v^2  X_t'(y_i)X_t'(y_j) f(y_i-y_j) dt + O(dt^{3/2})\;.
\end{align}
Averaging over the noise we obtain
\begin{multline}
 \frac{d\overline{g_t}}{dt} = -v \sum_{j=1}^n \overline{\partial_j G\big(X_t(y_1),\dots,X_t(y_n) \big) X'_t(y_j) }+\frac{v^2 f(0)}{2}\sum_{j=1}^n \overline{ \partial_j G\big(X_t(y_1), \ldots, X_t(y_n)\big) X_t''(y_j)} + \\
 +\frac{v^2}{2} \sum_{j,m=1}^n f(y_j - y_{m}) \overline{ \partial_j \partial_m G \big(X_t(y_1), \ldots, X_t(y_n)\big) X_t'(y_j) X_t'(y_{m}) }
\end{multline}
From the chain rule for the derivation with respect to the variables $\{ y_i\}$ it is easy to check that 
\begin{equation}
 \partial_{y_j}\partial_{y_m} g_t =X'_t(y_{j}) X'_t(y_{m}) \partial_j \partial_m G  + \delta_{j,m} X''_t(y_{j}) \partial_j G 
\end{equation}
which finally leads to:
\begin{equation}
\label{eq:dgeq}
\frac{d\overline{g_t}}{dt} = \Big[-v \sum_{j=1}^n\partial_{y_j}+ \frac{v^2}{2}\sum_{j,m=1}^n f(y_j - y_{m}) \partial_{y_{j}} \partial_{y_{m}} \Big] \overline{g_t} \;.
\end{equation}
It is useful to re-express \eqref{eq:dgeq} in terms of the Fokker-Planck Hamiltonian (and its hermitian adjoint). In order to do so, we introduce the operators $q_j$ and $p_j$, defined by their action on any smooth function $\omega(\bold{y})$ as $q_j \cdot \omega(\bold{y}) = y_j \omega(\bold{y}) $ and ${p}_j \cdot \omega(\bold{y}) = -\imath \partial_j \omega(\bold{y}) $. For these conjugate variables the canonical quantization holds $[q_i,p_j]=\imath \delta_{i,j}$. We can then define
\begin{align}
   & \mathcal{H}_{\rm FP} \equiv -\mathrm{i}v\sum_{i=1}^n p_i+ \frac{v^2}{2} \sum_{ij} p_i p_j  f(q_i -  q_j)\;, \label{eq:HFPdef} 
\end{align}
so that we can rewrite
\begin{equation}
\frac{d\overline{g_t}}{dt} = - \mathcal{H}_{\rm FP}^\dag \cdot \overline{g_t}
       \;, \qquad \mathcal{H}_{\rm FP}^\dag \equiv \mathrm{i}v\sum_{i=1}^n p_i+ \frac{v^2}{2} \sum_{ij} f( q_i -  q_j) p_i p_j  \;.  \label{eq:HFPTdef}
\end{equation}

To deduce the Fokker-Planck equation for $P(\xb | \yb)$, defined in the text, we need one more step. From the definition of $P(\xb | \yb)$, we have
\begin{equation}
\overline{g_t(\yb)} = \int d\bold{x}' \; G(\bold{x}') P_t(\bold{x}'| \yb)
\end{equation}
Since $G$ is an arbitrary function we choose $G({\bf x}')=\delta(\bold{x}'-\xb)$ to recover $\overline{g_t(\yb)} \rightarrow P(\xb | \yb)$ the jpdf for the initial points $\xb=(x_1,\dots , x_n)$ of the stochastic trajectories and finally deduce from \eqref{eq:HFPTdef}
\begin{equation}
\partial_t P_t(\xb|\yb) = - \mathcal{H}_{\rm FP}^T[\yb] \cdot P_t(\xb|\yb)
\end{equation}
This equation can be formally solved starting from the initial condition at $t = 0$ is $P_{t=0}(\xb | \yb) = \delta(\xb - \yb)$. It is useful to employ the bra $\bra{\bold{y}}$ and ket $\ket{\bold{x}}$ notation for the eigenstates of the position operators $\hat{q}_j$. Then, we can represent the probability distribution as:
\begin{equation}
 P_t(\xb | \yb) \equiv \braket{\yb | e^{-t \mathcal{H}_{\rm FP}^\dag} | \xb} = \braket{\xb | e^{-t \mathcal{H}_{\rm FP}} | \yb}
 \end{equation}
where the last equality follows from the hermitian conjugation. Therefore, the following equation must also holds:
 \begin{equation}
\partial_t P_t({\bf x} | {\bf y} ) = -\mathcal{H}_{\rm FP} [\xb]\cdot P_t(\xb | \yb) 
 \end{equation}
 where the action of the differential operator $\mathcal{H}_{\rm FP}[\xb]$ is now over the variables $\xb$. 
 More explicitly using \eqref{eq:HFPdef}, we arrive at
\begin{equation} \label{eq:FP1} 
\partial_t P_t(\xb | \yb) = v \left(\sum_{i=1}^n \partial_{i} + \frac{v}{2}\sum_{i,j=1}^n \partial_{i} \partial_{j} f(x_i - x_j)\right) P_t(\xb | \yb)  \;.
\end{equation}
which coincides with Eq.~\eqref{eq:PFP} given in the main text.

\section{Joint evolution of $r$ and $\kappa$}

In this section we give two equivalent methods to derive the joint evolution of the variables $r$ and $\kappa$ defined in the text.
The first one uses the Fokker-Planck equation and the second one is a direct derivation using stochastic equations. 

\subsection{Derivation of the evolution equation for $P_t(r, x_1', x_2')$}

We will first derive the evolution equation for the jpdf $P_t(x_1, x_2, x_1', x_2')$ of the random variables $x_1,x_2,x'_1,x'_2$ defined in the text. The starting point is the application of the Fokker-Planck equation \eqref{eq:FP1}
in the case of 4 trajectories $(x_1,x_2,x_3,x_4)$. 
To fix the variables $y_i$, we remind that $y_1$ and $y_2$ are kept fixed while two additional variables are taken infinitesimally away from $y_1$ and $y_2$. More explicitly
\begin{equation}
\label{eq:inicoord}
  x_1=X^{\pm}_t(y_1),\ \  x_2=X^{\pm}_t(y_2), \ \
  x_3=X^{\pm}_t(y_1+\epsilon),\ \  x_4=X^{\pm}_t(y_2+\epsilon) 
\end{equation}
We then perform the change of variables $x_1,x_2,x_1'=(x_3-x_1)/\epsilon, x_2'=(x_4-x_2)/\epsilon$ together with the limit= $\epsilon \rightarrow 0$. By applying this change of variables in the Fokker-Planck equation above and taking the limit $\epsilon \to 0$, one obtains the following equation for $P_t=P_t(x_1,x_2,x_1',x_2')$:
 \bea
\partial_t P_t  = v \bigg( (\partial_{x_1}+\partial_{x_2}) + \frac{v}{2} f(0) (\partial_{x_1}^2 + \partial_{x_2}^2) +  v\partial_{x_1} \partial_{x_2} f(x_1-x_2) - \frac{v}{2} F''(0) (\partial_{x_1'}^2 (x'_1)^2 + \partial_{x_2'}^2 (x'_2)^2 ) \nonumber \\
+ v\partial_{x_2} \partial_{x'_1} x_1' f'(x_1-x_2) - v\partial_{x_1} \partial_{x'_2} x_2' f'(x_1-x_2) 
- v\partial_{x'_1} \partial_{x'_2} x'_1 x'_2 f''(x_1-x_2) \bigg) P_t
\eea 
Then, we exploit the invariance under translation by making the change of variables from $(x_1,x_2,x_1',x_2')$ to
$(R=\frac{x_1+x_2}{2},r=(x_1-x_2)/\ell,x'_1,x'_2)$. Finally, integrating out the center of mass variable $R$, we obtain 
\begin{equation}
\label{eq:4point}
\partial_t  P  =v^2  \bigg(
 \frac{1}{\ell^2}\partial_r^2 (f(0)-f(r \ell)) - \frac{f''(0)}{2}(\partial_{x_1'}^2 (x'_2)^2 + \partial_{x_1'}^2 (x'_2)^2 ) - f''(r \ell) \partial_{x'_1} \partial_{x'_2} x'_1 x'_2  -
 \frac{1}{\ell}\partial_{r}  f'(r \ell) (\partial_{x'_1} x'_1 + \partial_{x'_2} x'_2 ) 
 \bigg) P  
 \end{equation}
\\

From this result we now obtain the Fokker-Planck equation for the joint distribution $P_t(\kappa,r)$ of the variable $\kappa$ and the variable $r$ defined as
\begin{equation}
\label{eq:rhosrdef}
    P_t(\kappa,r) \equiv \int dx_1' dx_2' \; P(r, x_1', x_2') \delta\left(\kappa -  \ln \Bigl( \frac{r^2}{x_1' x_2'} \Bigr)\right) \;.
\end{equation}
Interestingly, $P_t(\kappa, r)$ satisfies a closed evolution equation. To obtain it, we compute the time derivative of \eqref{eq:rhosrdef} using \eqref{eq:4point}, then we integrate by part and obtain
\bea
\label{eq:Prkappa}
\frac{\ell^2}{v^2} \partial_t P = \bigg[ \partial_r^2 (f(0)-f(\ell r)) + 4 \frac{f(0) - f(\ell r) + \ell r f'(\ell r)}{r^2} \partial_\kappa^2 
+ \ell^2 f''(0) (\partial_\kappa - \partial_\kappa^2) - \ell^2 f''(\ell r) \partial_\kappa^2 \nonumber \\
+ 2 \ell \partial_r \partial_\kappa f'(\ell r) + 4 \partial_r (f(0)-f(\ell r)) \frac{1}{r} \partial_\kappa 
+ 2  \frac{f(0)-f(\ell r)}{r^2}  \partial_{\kappa} \bigg] P\;.
\eea 
Then, it is easy to see that the FP equation in Eq.~\eqref{eq:Prkappa} is equivalent to the stochastic equations for $r$ and $\kappa$ (in Ito convention) given in \eqref{stoch1} of the main text. 

\subsection{Direct derivation of the stochastic equations for $r$ and $\kappa$}

Instead of working with the Fokker-Planck equation (4), we consider now an equivalent system of stochastic equations.
It is equivalent in the sense that the Fokker-Planck equation associated with this system coincides with (4). 
It provides an alternative way to derive Eq.~\eqref{stoch1} in the main text.
This stochastic system involves the set of trajectories $x_i(t)$, with $dx_i(t)=x_i(t+dt) - x_i(t)$, and reads
\be 
dx_i(t) = \mp v dt +  dW_i(t) \quad , \quad \left.\overline{dW_i(t) dW_j(t)}\right|_{t} = f(x_i(t)-x_j(t)) dt 
\ee
where the $W_i(t)$'s are mutually correlated Wiener processes in time and shall not be confused with $\WW_t(y)$ introduced in \eqref{eq:dWdef}. Formally, the expectation value $\left.\overline{O}\right|_t$ used here is conditioned to the position of the trajectories $x_i(t)$ at time $t$. The $\mp$ refers to $\pm$ chiralities, but the drift term plays no role in the following, so will be omitted in the following. Here, the initial condition $x_i(t=0) = y_i$ is assumed. Note that this system of equation can be also seen as the time-reversal of Eq.~\eqref{eq:langevin}.

Let us set $\ell=1$ and restore $\ell$ later. One performs the linear change of variable 
$x_3=x_1+\epsilon x'_1$, $x_4=x_2+\epsilon x'_2$ and $x_1-x_2=r$, which leads to (we keep the time dependence implicit for convenience)
\be 
dr = dW_1 -dW_2 \quad , \quad dx'_1=\frac{1}{\epsilon} (dW_3-dW_1) 
\quad , \quad dx'_2=\frac{1}{\epsilon} (dW_4-dW_2) 
\ee 
The correlations can be expressed in terms of $r,x'_1,x'_2$ (the center of mass decouples)
and performing the limit $\epsilon \to 0$ one finds with $j=1,2$
\be  \label{corr20} 
\overline{dr^2} = 2 (f(0)-f(r)) dt \quad , \quad \overline{dr dx'_j}= - f'(r) x'_j dt \quad , \quad \overline{(dx'_j)^2} = - f''(0) (x'_{j})^2 dt \quad , \quad 
\overline{dx'_1 dx'_2}= - f''(r) x'_1 x'_2 dt
\ee 
Now we have for $\kappa=\log(r^2/x'_1 x'_2)$ using Ito's rule
\be 
d \kappa = 2 \frac{dr}{r} - \frac{dx'_1}{x'_1} - \frac{dx'_2}{x'_2}  + g(r) dt 
\ee
where $g(r) dt$ is the Ito drift which can be computed using \eqref{corr20}
\be 
- \frac{1}{r^2} \overline{dr^2} + \frac{1}{2} \frac{\overline{(dx'_1)^2}}{(x'_1)^2}
+ \frac{1}{2} \frac{\overline{(dx'_2)^2}}{(x'_2)^2} = g(r) dt  \quad , \quad g(r) = \frac{2}{r^2} (f(0)-f(r)) - f''(0) 
\ee 
and depends only on $r$. We also need the correlations of the noise part
\be 
\overline{d\kappa dr} = 
2 \frac{dr^2}{r} - \frac{\overline{dr dx'_1}}{x'_1} - \frac{\overline{dr dx'_2}}{x'_2}
= B(r) dt \quad , \quad B(r)= \frac{4}{r} (f(0)-f(r)) + 2 f'(r) 
\ee 
\be 
\overline{d\kappa^2} = 4 \frac{\overline{dr^2}}{r^2}  +  \frac{\overline{(dx'_1)^2}}{(x'_1)^2} 
+  \frac{\overline{(dx'_2)^2}}{(x'_2)^2} - 4 \frac{\overline{dr dx'_1}}{r x'_1} 
- 4 \frac{\overline{dr dx'_2}}{r x'_2} + 2 \frac{\overline{dx'_1 dx'_2}}{x'_1 x'_2}  
= 2 C(r) dt \quad , \quad C(r)= 4 \frac{f(0) - f(r) + r f'(r)}{r^2} - f''(0) - f''(r)
\ee 
to which we must add $\overline{dr^2}= 2 A(r)$ with $A(r)=f(0)-f(r)$. 
Restoring $\ell$ one finds the equations in the text. 

\section{Numerical solution of Eq.~\eqref{stoch1}}
In this section, we provide some details about how to numerically solve the stochastic equations for $\kappa$ and $r$. The main difficulty arises because $r$ can become typically very small (see Eq.~\eqref{eq:udistrib}). It is thus useful to rewrite such a system of SDE in terms of another variable: 
\begin{equation}
    \rho = \ln r \; , \qquad r = e^\rho
\end{equation}
Then, using Ito's lemma, the stochastic equation for $\rho$ takes the form
\begin{equation}
    d \rho =  e^{-\rho} v dW_1 -  e^{-2 \rho} v^2 A(r(\rho)) dt 
\end{equation}
To solve the equation, we then discretize time $dt \to \Delta t$ and define 
\begin{equation}
\Delta W_1 = \sqrt{2 A(r)} \beta_1   \;, \qquad
\Delta W_2 = \sqrt{\frac{B(r)^2 }{2 A(r)}} \beta_1  + \sqrt{\left(2 C(r) - \frac{B(r)^2}{2 A(r)}\right)} \beta_2
\end{equation}
where $\beta_1, \beta_2$ are independently gaussian random variables with zero average and $\Delta t$ variance. To rewrite the equation in a more numerically stable way, we extract from the quantities $A(r), B(r)$, their leading behavior at small $r$
\begin{equation}
    \tilde{A}(r) = A(r)/r^2 \;, \quad \tilde{B}(r) = B(r) / r^3 
    \;. 
\end{equation}
so that $\tilde{A}(r), \tilde{B}(r)$ are both finite in the limit $r \to 0$. So we finally have the discrete evolution equations
\begin{align}
    &\rho(t + \Delta t) = \rho(t) + v \sqrt{2 \tilde{A}(r) \Delta t} \beta_1  - v^2 \tilde{A}(r) \Delta t \\
    &\kappa(t + \Delta t) = \kappa(t) +  v^2 g(r) \Delta t + v r^2 \sqrt{\frac{\tilde{B}(r)^2 }{2 \tilde{A}(r)} \Delta t} \beta_1  + \sqrt{\left(2 C(r) - \frac{r^4 \tilde{B}(r)^2}{2 
    \tilde{A}(r)}\right) \Delta t} \beta_2
\end{align}

\section{Analysis of the short distance regime for $\kappa$}
In the limit $\ell \ll 1$, one has Eq.~\eqref{eq:stoch2} in the main text that we report for convenience
\bea 
&& dr = r v dB_1 \quad , \quad d\kappa = - \frac{\ell^2}{6} r^2 v ( - \frac{v}{2} f^{(4)}(0) dt + dB_2) \label{eq:rkappasmallellSI}\\
&& \overline{dB_1 dB_1} = - f''(0) dt \quad , \quad \overline{ dB_2 dB_2} = - f^{(6)}(0) dt \quad , \quad \overline{ dB_1 dB_2 } = - f^{(4)}(0)  dt \label{eq:BBcorr}
\eea 
We will now show from these equations, that $\kappa$ satisfies a closed SDE which leads to the stationary measure given in the main text. We first of all solve the equation for the variable $r$, which takes the form
\bea 
r(t) = \exp\left[v B_1(t) + \frac{1}{2} v^2 f''(0) t\right]
\eea 
Injecting this solution in the equation for $\kappa$ and integrating in time we arrive at
\begin{equation}
\label{eq:kappastocint}
\kappa(t) = - \frac{\ell^2}{6} \int_0^t e^{2 v B_1(s) +  v^2 f''(0) s} v \left( - \frac{v}{2} f^{(4)}(0) ds + dB_2(s)\right)
\end{equation}
This equation gives already a closed representation for $\kappa(t)$. 

We can further simplify Eq.~\eqref{eq:kappastocint}
by making use of the time-reversal symmetry.

We denote $s = t - s'$ and introduce new processes as $d\tilde{B}_i(s') = dB_i( t - s')$. Therefore for both $i = 1,2$, we have
\begin{equation}
    B_i(t-s') = \int_{0}^{t-s'} dB_i(s'') = 
    \int_{0}^{t} dB_i(s'') +  \int_{t}^{t-s'} dB_i(s'') = B_i(t) - \int_0^{s'} d \tilde{B}_i(s'') = \tilde {B}_i(t) - \tilde{B}_i(s') 
\end{equation}
where $\tilde B_{i}(s)$ are equivalent Wiener processes which move in the opposite direction. We want now to rewrite Eq.~\eqref{eq:kappastocint} in terms of the reversed processes $\tilde{B}_i$. A little bit of care is needed for the stochastic integral. Indeed, writing explicitly the Ito's integral, we have
\begin{multline}
    \int_0^t e^{2 v B_1(s) +  v^2 f''(0) s} dB_2(s) = \lim_{n\to\infty} \sum_i e^{2 v B_1(s_i) +  v^2 f''(0) s_i} (B_2 (s_{i+1}) - B_2(s_i)) =\\= e^{2 v \tilde{B}_1(t) +  v^2 f''(0) t}\lim_{n\to\infty}  \sum_j e^{-2 v \tilde{B}_1(\tilde{s}_j) -  v^2 f''(0) \tilde{s}_{j}} (\tilde{B}_2 (\tilde{s}_{j}) - \tilde{B}_2(\tilde{s}_{j-1}))
\end{multline}
where the $s_i$'s are a partition of $[0,t]$ and $\tilde{s}_{j} = t - s_i$ defines an equivalent partition, with $j = N - i$. Clearly, the last expression does not converge to a stochastic integral in the Ito form. We thus expand the exponent arriving at
\begin{equation}
    \int_0^t e^{2 v B_1(s) +  v^2 f''(0) s} dB_2(s)=
    e^{2 v \tilde{B}_1(t) +  v^2 f''(0) t} \left[
    \int_0^t e^{- 2 v B_1(s)-  v^2 f''(0) s} d\tilde{B}_2(s) + 
    2 v f^{(4)}(0) 
    \int_0^t e^{- 2 v B_1(s)-  v^2 f''(0) s} ds \right]
\end{equation}

Applying these transformations to \eqref{eq:kappastocint} we arrive at
\begin{equation}
\label{eq:kappastocintrev}
\kappa(t) = \underbrace{- \frac{\ell^2}{6} e^{2 v \tilde{B}_1(t) +  v^2 f''(0) t}}_{\kappa_1} \underbrace{\int_0^t e^{- 2 v \tilde{B}_1(s) -  v^2 f''(0) s} v \left(  \frac{3v}{2} f^{(4)}(0) ds + d\tilde{B}_2(s)\right)}_{\kappa_2}
\end{equation}

Using Ito's lemma, $d\kappa = \kappa_2 d\kappa_1 + \kappa_1 d\kappa_2 + d\kappa_1 d\kappa_2$, which, after collecting different contributions, leads to
\begin{equation}
\label{eq:kappasingleeq}
    d\kappa =  
    v^2 \left(\frac{ \ell^2}{12}  f^{(4)}(0) - f''(0) \kappa\right)dt -\frac{ \ell^2 v }{6} d\tilde{B}_2(t)
    + 2 v d\tilde{B}_1(t) \kappa   
\end{equation}
The correlations of the $d\tilde{B}_j(t)$ are the same as the ones of the $dB_j(t)$ in Eq.~\eqref{eq:BBcorr}.
However, $\kappa(t)$ defined by Eq. \eqref{eq:kappastocintrev} is a {\it different process} in $t$ 
than $\kappa(t)$ defined by Eq.~\eqref{eq:kappastocint}. 
At fixed $t$ the two random variables have the same law, but
as $t$ is varied the trajectories are different (since the relation between $\tilde B_i$ and $B_i$ involves $t$ explicitly).
As a consequence, the two stochastic equations 
\eqref{eq:kappasingleeq} and \eqref{eq:kappastocint} are inequivalent although they lead to the same single-time distribution for $\kappa(t)$.
One illustration of that is that while the second process converges, i.e. $\kappa(t) \to \kappa_\infty$, where the distribution of
$\kappa_\infty$ is given below, the first process is ergodic (with the same law) as we show below. 

We can recast \eqref{eq:kappasingleeq}
as an equation with a single Brownian process $dB$ (we are using that $a_1 dB_1 + a_2 dB_2 = \sqrt{- a_1^2 f''(0) - a_2^2 f^{(4)}(0) - 2 a_1 a_2 f^{(6)}(0)} d\tilde{B}$, where $d\tilde{B}$ is a new Wiener process with standard normalization, $d\tilde{B}^2 = dt$)
\begin{equation}
d\kappa =  
    v^2 \left(\frac{ \ell^2}{12}  f^{(4)}(0) - f''(0) \kappa\right)dt 
    + \frac{v d\tilde{B}}{6} \sqrt{-  \left(\ell^4 f^{(6)}(0)-24 \ell^2 f^{(4)}(0) \kappa +144 f''(0) \kappa ^2\right)}
\end{equation}
With the change of variable $\kappa = \kappa_0 (\omega / \omega_0 - 1)$ where we defined as in the main text \eqref{eq:kappatoY}
\begin{equation}
    \kappa_0= - \frac{\ell^2}{12} \frac{f^{(4)}(0)}{f''(0)} \;, \qquad \omega_0  = \frac{1}{\sqrt{\frac{f^{(6)}(0) f''(0)}{f^{(4)}(0)^2}-1}}
\end{equation}
Note that $\omega_0$ is real and positive as it is guaranteed by the positivity of the Fourier transform $\hat{f}(k)>0$ of $f(x)$. Indeed,
\bea
 f^{(6)}(0) f''(0)-f^{(4)}(0)^2 = \int k^2 \hat f(k) \int k^6 \hat f(k) - \left(\int k^4 \hat f(k)\right)^2 > 0 
\eea 
which is a consequence of the Cauchy-Schwartz inequality.

The SDE for $\omega$ becomes
\begin{equation} \label{eq:stochomega} 
    d\omega = 2 \theta \omega dt  + \sqrt{8 \theta} \sqrt{1 + \omega^2}\; dW
\end{equation}
Accounting for the initial condition $\omega(t = 0) = \omega_0$, the solution of this SDE takes the form 
\begin{equation} \label{solu2} 
    \omega(t) = e^{\sqrt{8\theta} B(t) - 2 \theta t}\left(\omega_0 +  \sqrt{8\theta}\int_0^t e^{-\sqrt{8 \theta}B(s) + 2 \theta s} d \gamma_s\right)
\end{equation}
which is then a Bougerol variable with drifted Brownian motion in the exponent \cite{vakeroudis2012bougerols}. It is useful to do the change of variable~\eqref{eq:stochomega} 
\be 
\omega = \sinh Y \quad , \quad Y= Y(\omega) = {\rm argsinh} \omega 
\ee 
One has
\be 
Y'(\omega) = \frac{1}{\sqrt{1+ \omega^2} } = \frac{1}{\cosh Y} \quad , \quad 
Y''(\omega) = - \frac{\omega}{(1+ \omega^2)^{3/2}} 
\ee 
and from Ito's rule, it follows Eq.~\eqref{eq:Ystoc} in the main text.

\section{Stationary measure for $\kappa$ for any finite $\ell$}

\subsection{Backward method}
We now apply the backward method to the full stochastic equation for $r(t),\kappa(t)$  in \eqref{stoch1}. We define 
\be 
\label{eq:Qkdef}
Q_k(r_0,t) \equiv \overline{ e^{- i k  \kappa(t) }}^{r_0} 
\ee 
where the superscript $r_0 = r(t=0)$ indicates the initial condition for the variable $r$. In the end we will set $r_0 = 1$ as it is required in our case, but it is useful to keep it free. One has 
\bea
&& Q_k(r_0,t+dt) = \overline{ e^{- i k (v^2 g(r_0) dt + v \, dW_2(0) ) } Q(r_0 + v dW_1(0),t)}^{r_0} 
\eea 
expanding with Ito's lemma and averaging we arrive at
\be 
\label{eq:Qpfull}
\partial_t Q_k = v^2 \left( 
A_\ell(r) \partial_r^2 -i k  B_\ell(r) \partial_r  - k^2  C_\ell(r) - i k g_\ell(r) \right) Q_k(r,t)  \;.
\ee 
with the boundary conditions
\be \label{eq:Qtboundary} 
Q_k(r_0,0)=1 \quad , \quad Q_k(r_0=0,t) = 1  \quad , \quad  \lim_{r_0 \to +\infty} Q_k(r_0,t)  =
 e^{- 2 \theta (k^2 + i k) t}
\ee 
and we recall that
\bea \label{eq:ABC2} 
&& A_\ell(r)= \frac{f(0) - f(\ell r)}{\ell^2} ~,~ B_\ell(r)= 2  \frac{f'(\ell r)}{\ell}  + 4 \frac{f(0)-f(\ell r)}{\ell^2 r}   \nonumber \\
&& C_\ell(r)=  4 \frac{f(0) - f(\ell r) + \ell r f'(\ell r)}{\ell^2 r^2} - f''(0) - f''(\ell r)  \nonumber \\
&& g_\ell(r)= - f''(0)-2 \frac{f(0)-f(\ell r)}{\ell^2 r^2} 
\eea 
The second condition in \eqref{eq:Qtboundary} comes from the fact that $d\kappa=0$ and $dr=0$ for $r=0$ since $g(0)=0$ and $C(0)=0$. The third condition is obtained using the fact that the dynamics of $\kappa$ is pure diffusion at large $r_0$.

Let us denote $Q_k(r)$ the stationary solution of \eqref{eq:Qpfull}.
It thus satisfies
\be \label{eq:Qeq}
A_\ell(r) Q_k''(r) -i k B_\ell(r) Q_k'(r)- (k^2 C_\ell(r) + i k g_\ell(r)) Q_p(r) = 0 
\ee 
with the boundary conditions 
\be 
\label{eq:boundQ}
Q_{k}(r_0=0) = 1  \quad , \quad Q_{k}(r_0 \to + \infty) = 0
\ee 
From this stationary solution one obtains the stationary measure $\Pstat(\kappa)$ for $\kappa$ by Fourier inversion
\be 
\label{eq:PkappaQ}
\Pstat(\kappa) = \int_{-\infty}^{+\infty} \frac{dk}{2 \pi} e^{i k \kappa} Q_k(r=1)
\ee 

\subsection{Schrodinger equation for the stationary measure}

We can further simplify Eqs.~\eqref{eq:Qtboundary} and \eqref{eq:Qeq} by removing the first derivative term. This can be achieved setting
\be
\label{eq:Qfact}
Q_k(r) = \phi_k(r) G_k(r) \;.
\ee 
If one chooses $\phi_k(r)$ so that
\be 
\frac{\phi_k'(r)}{\phi_k(r)} = i k \frac{B(r)}{2 A(r)} \quad , \quad \phi_k(r)= e^{ i k \int_0^r dr' \frac{B(r')}{2 A(r')} } = \left(\frac{\ell^2 r^2 f''(0)}{2 (f(\ell r) - f(0))}\right)^{i k}
\ee 

then one finds that $G_p(r)$ satisfies the Schrodinger equation 
\be
\label{eq:GSchro1}
  - G_k''(r) - k (k + i) V(r) G_k(r) = 0 
\ee
with the potential 
\be
V(r) = \frac{{\ell}^2 \left(f'({\ell}
   r)^2+(f(0)-f({\ell} r)) \left(f''({\ell}
   r)+f''(0)\right)\right)}{(f(0)-f({\ell} r))^2} = - \frac{d^2}{dr^2} \log[f(0)-f(\ell r)] + \frac{\ell^2 f''(0)}{f(0)-f(\ell r) }
\ee
on the positive half-space $r \geq 0$ with the boundary condition 
\be 
\label{eq:Gboundary}
G_k(0) = 1 \;, \qquad \lim_{r \to \infty} G_k(r) = 0 \;.
\ee

Before solving various cases, let us indicate a nice symmetry property. One notes that the 
dependence in $k$ in \eqref{eq:GSchro1} is only via the prefactor $\gamma \equiv k(k+i)$ of the potential: $G_k(r) \equiv \mathcal{G}(\gamma; r)$. This implies that
\be 
\int d\kappa \Pstat(\kappa) e^{- i k \kappa} = \phi_k(1) G_k(1)=  e^{i k \kappa_0} \mathcal{G}(\gamma; 1)
\ee 
where $\kappa_0$ is explicitly
\be 
\kappa_0 =  \frac{1}{ik}\log \phi_k (1) = - \log\Bigl( \frac{2 (f(\ell) - f(0))}{\ell^2  f''(0)} \Bigr)
\ee
Hence, from Eq.~\eqref{eq:PkappaQ}
\be
\Pstat(\kappa) = \int \frac{dk}{2 \pi} e^{ i k (\kappa + \kappa_0)  } \mathcal{G}(k(k+i);1) = e^{\frac{\kappa+\kappa_0}{2}} \int \frac{du}{2 \pi} e^{ i u (\kappa + \kappa_0)  } \mathcal{G}(u^2 + \frac{1}{4};1) 
\ee
where in the last equality we changed variable $k = u - i/2$. 
It implies in particular that 
\be 
\Pstat(\kappa) = e^{\frac{\kappa+\kappa_0}{2}} \tilde P( |\kappa + \kappa_0| ) \quad , \quad \tilde P(y)= \int \frac{du}{2 \pi} e^{ i u y } \mathcal{G}(u^2 + \frac{1}{4};1) 
\ee 
or equivalently the symmetry (reminiscent of a Nishimori condition, or a Galavotti-Cohen theorem) 
\be 
\frac{\Pstat(-\kappa_0 + \beta)}{\Pstat(-\kappa_0 -\beta) } = e^{\beta}
\ee

\subsection{Proof of the right $3/2$--tail}
Here, we show that for a smooth noise correlation $f(x)$, the right tail of the stationary distribution is always a power-law $\Pstat(\kappa)$
$ \propto
\kappa^{-3/2}$ for $\kappa \to +\infty$,
independently of $\ell$.
Thanks to \eqref{eq:Qkdef}, it is enough to prove the following expansion at small $k$ for its Fourier transform
\begin{equation}
\label{eq:Qsmallk}
    Q_k(r=1) = 1 + C \sqrt{k} + O(k).
\end{equation}
with $C$ is a constant (see below). 
To prove Eq.~\eqref{eq:Qsmallk}, we proceed as follow. First of all, since $\phi_k(r)$ is analytic in $k$, using \eqref{eq:Qfact}, we can focus on the small $k$ expansion of $G_k(r)$. The small $k$ behavior of the solution of Eq.~\eqref{eq:GSchro1} can be obtained setting $x = r \sqrt{\gamma}$, with $\gamma = k (k + i)$. Then, setting $G_k(r) = g(r \sqrt{\gamma})$, we can rewrite \eqref{eq:GSchro1} in the limit $k \to 0$ as
\begin{equation}
    - g'(x) - V_\infty g(x) = 0 \quad \Rightarrow \quad g(x) = e^{- \sqrt{-V_\infty} x} 
\end{equation}
where we set $V_\infty = \lim_{r \to \infty} V(r)$ and enforced the boundary conditions \eqref{eq:Gboundary}. This implies
\begin{equation}
    G_k(r) \sim g(r \sqrt{\gamma}) = e^{-\sqrt{-V_\infty \gamma} r} \;, \quad \forall r = O(\gamma^{-1/2})
\end{equation}
This is still not enough because we require the expansion \eqref{eq:Qsmallk} for $r = 1$. However, fixing $\delta>0$ and $r$, we can write
\begin{equation}
    | G'_k(r) - G'_k(\delta / \sqrt{\gamma}) | = \left| \int_{\delta/\sqrt{\gamma}}^r dr' G_k''(r')\right| \leq \gamma \int_{\delta/\sqrt{\gamma}}^r dr' |V(r') G_k(r')| \leq K \gamma (r - \delta/ \sqrt{\gamma})
\end{equation}
where we set $K = \sup_r |V(r) G_k(r)|$, which is finite for a sufficiently smooth $f(x) \in C^6$. As a consequence, at small $\gamma$,
\begin{equation}
G'_k(r) = G'_k(\delta / \sqrt{\gamma}) + O(\delta \sqrt{\gamma}) = -\sqrt{-V_\infty \gamma}\; (g(\delta) + O(\delta)) \stackrel{\delta \to 0}{\longrightarrow} -\sqrt{-V_\infty \gamma} + O(\gamma) 
\end{equation}
Finally, integrating over $r$
\begin{equation}
    G_k(1) = 1 + \int_0^1 dr \; G_k'(r) = 1 - \sqrt{-V_\infty \gamma} + O(\gamma)
\end{equation}
which proves Eq.~\eqref{eq:Qsmallk} with $C = \sqrt{-V_\infty}$.

\subsection{Small $\ell$ limit} 

At small $\ell$ one finds that the potential is a harmonic oscillator
\bea
\label{eq:harmV}
V(r) = 
\frac{{\ell}^4 r^2 \left(f^{(4)}(0)^2 - f^{(6)}(0)
   f''(0) \right)}{36 f''(0)^2} + O(\ell^6 r^4) 
\eea
and 
\be
\kappa_0 = \tilde{\kappa}_0 \ell^2 + O(\ell^4) \quad , \quad \tilde{\kappa}_0 = -\frac{f^{(4)}(0)}{12   f''(0)}
\ee 

We see that $\kappa_0$ is $O(\ell^2)$ as $\ell \to 0$,
so the random variable $\kappa = O(\ell^2)$ in this limit. Therefore, we can obtain a $\ell$ independent limit by scaling $k = \tilde k/\ell^2$. In terms of this variable 
the potential term in the Schrodinger equation has a finite limit
\be t
- k (k+i) V(r) \simeq\frac{4 \tilde k^2 r^2 \tilde{\kappa}_0^2} {\omega_0^2}
\ee 
where we have defined as in the text 
\be 
\omega_0 = \frac{1}{\sqrt{ \frac{f''(0) f^{(6)}(0)}{f^{(4)}(0)^2}-1}} 
\ee
A solution of Eq.~\eqref{eq:GSchro} with the potential \eqref{eq:harmV} and the boundary conditions \eqref{eq:Gboundary} can be expressed in terms of Bessel function as
\be 
\label{eq:Gsolsmallell}
G_k(r)= 
\frac{2^{3/4} }{\Gamma(1/4)} \left(\frac{\tilde{\kappa}_0 |\tilde{k}|}{\omega_0}\right)^{1/4} \sqrt{r} K_{\frac{1}{4}}\left(\frac{|\tilde{k}| r^2 \tilde{\kappa}_0}{\omega_0} \right)
\ee 
and the prefactor has been fixed imposing that $G_k(r= 0) = 1$. This leads to
\be 
\label{eq:Qk1smallell}
Q_k(r=1)= \frac{2^{3/4} }{\Gamma(1/4)} \left(\frac{\tilde{\kappa}_0 |\tilde{k}|}{\omega_0}\right)^{1/4} 
e^{i \tilde{k} \tilde{\kappa}_0} K_{\frac{1}{4}}\left(\frac{|\tilde{k}| \tilde{\kappa}_0}{\omega_0} \right)
\ee 
which allows to determine $\Pstat(\kappa)$ by Fourier inversion from \eqref{eq:PkappaQ}. 

One can check that this coincides with the result in the text, identifying $\tilde{\kappa} = \tilde{\kappa}_0 \left( \frac{\omega}{\omega_0} -1 \right)$. Equivalently, we obtain the scaling form 
\be 
\Pstat(\kappa) \stackrel{\ell \ll 1}{\simeq} \frac{1}{\ell^2} \tilde P (\frac{\kappa}{\ell^2})  \;, \qquad 
\tilde{P}(\tilde\kappa) \equiv \frac{C \omega_0}{\tilde \kappa_0}  \left[1+ \omega_0^2 \Bigl(\frac{\tilde \kappa+ \tilde \kappa_0}{\tilde \kappa_0}\Bigr)^2 \right]^{-3/4}  
\ee 
and one can check that the Fourier transform of $\tilde P$
\be 
\int \frac{d\tilde k}{2 \pi} e^{ i \tilde k \tilde \kappa}  \tilde P(\tilde \kappa) = Q_k(r=1)
\ee 
as given in \eqref{eq:Qk1smallell}. This can be seen 
using the identity
\be 
\int dx e^{i k x} \frac{1}{(1+x^2)^{3/4}} = \sqrt{2 \pi} 
\frac{ (2 | k|)^{1/4}  K_{\frac{1}{4}}(|k|
   )}{\Gamma \left(\frac{3}{4}\right)}
\ee

\subsection{Large $\ell$ limit}
At large $\ell$, under the hypothesis that $f(x)$ and its derivatives decay at infinity, the potential term reaches a constant value
\be 
-k (k + i) V(r) \simeq -k (k + i) \ell^2 \frac{f''(0)}{f(0)}  \sim   \frac{ 2 i \tilde{k} \theta}{f(0)} \;.
\ee 
where we used once again the scaling $k = \tilde{k} / \ell^2$, which implies $\kappa = \ell^2 \tilde{\kappa}$, but with $\ell \to \infty$ in this case. From this potential, we immediately derive the solution respecting the boundary conditions \eqref{eq:Gboundary} in the form
\begin{equation}
    \label{eq:Gsollargeell}
    G_k(r) = e^{- \sqrt{2 i \tilde{k} \theta / f(0)} r}
\end{equation}
Note that at large $\ell$
\be
\kappa_0 \stackrel{\ell \to \infty}{=}\log( \ell^2 (-f''(0))/2 f(0))
\ee
so that $k \kappa_0 \stackrel{\ell \to \infty}{\longrightarrow} 0$ and therefore $Q_k(r) = G_k(r)$. Inverting the Fourier transform \eqref{eq:PkappaQ}, we obtain once again the stationary distribution
\begin{equation}
\Pstat(\kappa) \stackrel{\ell \gg 1}{\simeq} \frac{1}{\ell^2} \tilde p \bigl(\frac{\kappa}{\ell^2}\bigr)  \;, \qquad 
\tilde{p}(\tilde\kappa) \equiv \sqrt{\frac{\theta }{2\pi f(0)}}\frac{ e^{-\frac{\theta }{2 f(0) \tilde{\kappa}}}}{\tilde{\kappa}^{3/2}} \Theta(\tilde{\kappa})
\ee 
Equivalently, denoting $\kappa = \theta \ell^2 \xi / f(0)$ one finds that $\xi$ is distributed according to 
$\mathcal{L}(\xi)$ in Eq. \eqref{eq:Pstatlargeell} in the text, i.e. the stable one sided Levy distribution of index $1/2$.

\subsection{Solvable cases for $f(x)$} 
For some particular choice of the noise correlation function $f(x)$, the potential $V_k(r)$ takes a form which is explicitly integrable. In Table~\ref{tab:solvablef}, we list a few interesting cases. 
\begin{table}[t!]
    \centering
    \begin{tabular}{|c|c|c|c|c|}
    \hline
        $f(x)$  & $V(r)/\ell^2$ & $\hat f (k)$ & Smoothness\\
    \hline
        $e^{-|x|}$ & $\frac{1}{(1-e^{-r \ell})^2}$ &
        $\frac{2}{1 + k^2}$ & $C^0$\\                 $2e^{-|x|} - e^{-2 |x|}$ & $- \frac{2}{1 - e^{- \ell r}}$ & $\frac{12}{k^4+5 k^2+4}$ & $C^2$ \\
        $\frac{1}{\cosh x}$ & $- \tanh(\ell r)^2$ & $\frac{\pi}{\cosh(k \pi/2)}$ & $C^{\infty}$ \\
        $\frac{1}{(\cosh x)^2}$ & $- 2\tanh(\ell r)^2$ & $\frac{\pi k}{\sinh(k \pi/2)}$ & $C^{\infty}$ \\
        \hline
    \end{tabular}
    \caption{A few examples of noise correlation functions $f(x)$ leading to Schr\"odinger equations with a solvable potential $V(r)$.}
    \label{tab:solvablef}
\end{table}
Here, we focus on the case 
\begin{equation}
\label{eq:solvable}
f(x)=1/\cosh(x)
\end{equation}
which is analytic and fastly decaying. Setting $\gamma = k (k + i)$, the solution $G_{k}(r)$ respecting the boundary conditions \eqref{eq:Gboundary} can be expressed in terms of hypergeometric function
\begin{equation}
\label{eq:G2f1}
    G_k(r) = 
    e^{-\sqrt{\gamma } \ell r} (1+\tanh (\ell r))^{\sqrt{\gamma }} \, 
    \frac{\hyp\left(
    \frac{\sqrt{\gamma }}{2}-\frac{1}{2} \sqrt{\gamma +\frac{1}{4}}+\frac{1}{4},
    \frac{\sqrt{\gamma }}{2}+\frac{1}{2} \sqrt{\gamma +\frac{1}{4}}+\frac{1}{4} ;
    \sqrt{\gamma }+1;\frac{1}{\cosh(\ell r)^2}\right)}
    {\hyp\left(
    \frac{\sqrt{\gamma }}{2}-\frac{1}{2} \sqrt{\gamma +\frac{1}{4}}+\frac{1}{4},
    \frac{\sqrt{\gamma }}{2}+\frac{1}{2} \sqrt{\gamma +\frac{1}{4}}+\frac{1}{4} ;
    \sqrt{\gamma }+1;1\right)}
\end{equation}
Equivalently, this expression can be represented in terms of generalised Legendre functions
\be 
G_{k}(r)= \frac{P_{\frac{1}{2} \left(\sqrt{1 + 4\gamma}-1\right)}^{- \sqrt{\gamma}} (\tanh
   (\ell r))}{P_{\frac{1}{2} \left(\sqrt{1 + 4\gamma}-1\right)}^{- \sqrt{\gamma}}(0)}
\ee 
As a first check, we verify that the solution Eq.~\eqref{eq:G2f1} reproduces the known solutions in the small/large $\ell$ limits.

\subsubsection{Asymptotic limits $\ell \to \infty$}
At large $\ell$, we simply have
\begin{equation}
\label{eq:limitlargeellsolvable}
    \lim_{\ell \to \infty} Q_{k'/\ell^2} (r)  =    \lim_{\ell \to \infty} G_{k'/\ell^2} (r) = e^{- r \sqrt{i k'}}
\end{equation}
in agreement with \eqref{eq:Gsollargeell} ($\theta = 1/2$ for \eqref{eq:solvable}). The limit in \eqref{eq:limitlargeellsolvable}
can be easily obtained using \eqref{eq:G2f1} using that 
\be
\lim_{\gamma \to 0} \hyp\left(
    \frac{\sqrt{\gamma }}{2}-\frac{1}{2} \sqrt{\gamma +\frac{1}{4}}+\frac{1}{4},
    \frac{\sqrt{\gamma }}{2}+\frac{1}{2} \sqrt{\gamma +\frac{1}{4}}+\frac{1}{4} ;
    \sqrt{\gamma }+1;x\right) = \hyp(0,\frac12,1;x) = 1
\ee
irrespectively of $x$.
 
\subsubsection{Small $\ell$ check}
In this case, the limit is less trivial as $k = k'/\ell^2$ becomes large in the limit of small $\ell$. So that simultaneously the parameters of the hypergeometric are diverging, while its argument is going to $1$.
Thus, we first apply the transformation between hypergeometric functions
\begin{multline}
    \hyp(a,b,c; z) = \frac{(1-z)^{-a-b+c} \Gamma (c) \Gamma (a+b-c)\hyp(c-a,c-b;-a-b+c+1;1-z)}{\Gamma (a) \Gamma (b)}+\\+\frac{\Gamma (c) \Gamma (-a-b+c) \, \hyp(a,b;a+b-c+1;1-z)}{\Gamma (c-a) \Gamma (c-b)}
\end{multline}
Then, we use that
\begin{equation}
    \lim_{\ell \to 0} \hyp\left(\frac{1}{4} \left(2 \sqrt{\gamma }-\sqrt{1 + 4\gamma}+1\right),\frac{1}{4} \left(2 \sqrt{\gamma }+\sqrt{1 + 4\gamma}+1\right);\frac{1}{2};\tanh(\ell)^2\right) = 
    \frac{e^{k'/2} (k')^{1/4} \Gamma \left(\frac{3}{4}\right) I_{-\frac{1}{4}}(k'/2)}{\sqrt{2}}
\end{equation}
to recover, after some manipulations, Eq.~\eqref{eq:Qk1smallell}.
\section{Distribution of the stress-energy tensor}
As we saw in the main text, the distribution of the stress energy tensor can be deduced from the one of $\kappa$ in the limit of small $\ell$, simply using \eqref{eq:kappaTrel}. An alternative approach is to compute explicitly the time evolution of the stress energy tensor directly from the expression of the Hamiltonian. 

\subsection{Direct derivation}
Consider the explicit expression of the Hamiltonian \eqref{eq:genham} in terms of the stress energy tensor
\begin{equation}
 \hat{H} = v \int dx (1 + \eta(x,t)) (\hat{T}^+(x)+ \barT(x)) \;.
\end{equation}
To derive the time evolution of an operator, we introduce the infinitesimal generator of time evolution 
\begin{equation}
 d\hat{H} = v\int dx (dt + d\WW_t(x)) (\hat{T}^+(x)+ \barT(x))  \;.
\end{equation}
and define the time evolution $\mathcal{U}_t$ operator up to time $t$ with the equation
\begin{equation}
    \mathcal{U}_{t+dt} = e^{-i d\hat{H}} \mathcal{U}_{t}
\end{equation}
The evolution of an operator $\hat O(t) \equiv \mathcal{U}_t^{\dag} \hat O \mathcal{U}_t$ 
takes the form
\begin{equation}
 d\hat{O}(t) = \mathcal{U}_t^\dag e^{\imath d\hat{H}} \hat O e^{-\imath d\hat H} \mathcal{U}_t  - \hat O(t) = \mathcal{U}_t^\dag \left(
 \imath [d\hat{H}, \hat O] - \frac 12 [\hat{dH}, [\hat{dH}, \hat O]] + \dots \right) \mathcal{U}_t 
\end{equation}
Note that because of the Ito's convention we need to keep terms up to the double commutators. 

Consider the particular case of $\hat O = \hat T^\pm(y)$, the stress energy tensor at the point $y$. We have for the equal-time commutator
\begin{align}
[\hat{T}^+(x), \hat{T}^+(y)] &= - \imath (2 \delta'(x-y) \hat{T}^+(y)  - \delta(x-y) \hat{T}^{+ \prime}(y) - \frac{c}{24\pi} \delta'''(x-y))\\
[\barT(x), \barT(y)] &=  \imath (2 \delta'(x-y) \barT(y)  - \delta(x-y) \hat{T}^{- \prime}(y) - \frac{c}{24\pi} \delta'''(x-y))
\end{align}
From this we deduce
\begin{equation}
 [d\hat H, \hat T^{\pm}(y)] = \pm 2\imath  \hat T^{\pm}(y)v d\WW_t'(y) \pm \imath (dt + d\WW_t(y)) v \hat T^{\pm \prime}(y) \mp \imath \frac{c}{24\pi} v d\WW_t'''(y)
\end{equation}
where $d\WW_t'(x) = \partial_x d\WW_t(x)$ (the noise is smooth in space) and similarly for the higher derivatives. 
This implies the evolution equation for the operators $\hat T^\pm(y,t)$, which reads 
\begin{multline}
\label{eq:stocT}
  d\hat T^{\pm}(y,t) =   
   \mp (2 \hat T^{\pm}(y,t) v d\WW_t'(y) + (dt + d\WW_t(y)) v  \hat T^{\pm \prime}(y,t) - \frac{c}{24\pi} v d\WW_t'''(y)) + \\ + 
   \frac{c v^2 f^{(4)}(0)}{48 \pi}  dt
-  f''(0) v^2 \hat T^{\pm}(y,t) dt  + \frac{1}{2} f(0) v^2 \partial_y^2  \hat T^{\pm}(y,t) dt  \;. 
\end{multline}
Although the two chiral components do not 
couple at the CFT level, their evolutions are not statistically independent since they feel the same noise.{There are several quantities that one can study from there. One 
is the noise average $\overline{\hat{T}^{\pm}(y,t)}$, which is still a quantum operator. Its evolution is obtained by taking the noise average of
\eqref{eq:stocT}} and reads
\begin{eqnarray}
&& \partial_t  \overline{\hat T^{\pm}(y,t)}  = \frac{c v^2 f^{(4)}(0)}{48 \pi}
-  f''(0) v^2 \overline{ \hat T^{\pm}(y,t) } + \frac{1}{2} f(0) v^2  \partial_y^2  \overline{ \hat T^{\pm}(y,t) } \mp v \partial_y \overline{\hat T(y,t)}
\end{eqnarray} 

{Another observable is the quantum expectation $\langle \hat T^\pm(y,t) \rangle \equiv \braket{\Psi_0 |\hat T^\pm(y,t) | \Psi}$ on any translational invariant state $\ket{\Psi}$ 
in a given noise realisation. It satisfies a stochastic differential equations obtained by taking the quantum expectation of \eqref{eq:stocT} (which leads
to a similar equation as \eqref{eq:stocT} but now for a scalar $\langle \hat T^\pm(y,t) \rangle$). Note that it describes the coupled stochastic evolution of the two fields $\langle \hat T^\pm(y,t) \rangle$, a complicated problem.
Here we will only focus on (i) noise moments, which have a solvable dynamics (ii) the one-point PDF of $\langle \hat T^\pm(y, t) \rangle$. This one-point distribution
being independent of $y$ at $t=0$, it remains independent of $y$ for all times. In addition it does not depend on the chirality, thus we omit the $\pm$ superscript and denote $T(t) \equiv \langle \hat T^\pm(y, t) \rangle$ the corresponding random variable.}
Then, one can show that its PDF can be obtained from the stochastic equation \eqref{eq:dTsimple}
\be
\label{eq:dTsimple}
  d T  =   
   2 T v dB_1(t) - \frac{c}{24\pi} v dB_2(t) 
 +  \frac{c v^2 f^{(4)}(0)}{48 \pi} dt -  f''(0) v^2 T dt   
\ee
with $dB_1(t) = \mp dW'(y,t)$ and $dB_2(t) = \mp dW'''(y,t)$. This can be seen intuitively: indeed, we expect that the spatial derivative terms in \eqref{eq:stocT}
are irrelevant since the one point PDF of $T \equiv \langle \hat T^\pm(y, t) \rangle$ does not depend on $y$. More formally,
one can prove that \eqref{eq:dTsimple} and \eqref{eq:stocT} lead to the same evolution equation for the noise average $\overline{Z(T)}$ for any smooth function $Z$.
Although Eq. \eqref{eq:dTsimple} was obtained here by a completely different method, one can check that Eq.~\eqref{eq:dTsimple} is equivalent to \eqref{eq:kappasingleeq}
with the correspondence 
\begin{equation}
    T =  \lim_{\ell \to 0}\frac{c}{4 \pi \ell^2}  \kappa \;,
\end{equation} 
noting also that the noise satisfies  \eqref{eq:BBcorr}.

For the translationally invariant initial state chosen here, it is easy to obtain from \eqref{eq:dTsimple} the recursive equation for the moments (over the noise) of the quantum expectation 
of $T$ as 
\begin{equation}
\label{eq:momTrec}
    \partial_t \overline{T^n} = -\frac{v^2 c^2 f^{(6)}(0) n (n-1) }{1152 \pi ^2} \overline{T^{n-2}}+\frac{c v^2 f^{(4)}(0) (4n-3) n }{48 \pi }\overline{T^{n-1}} - n (2n-1) v^2 f''(0) \overline{T^n};
\end{equation}
These can be solved with $T(t=0)=0$, which corresponds to $|\Psi\rangle = |\Psi_0 \rangle$ being the ground state. This leads to the first two moments \begin{align}
\label{T1moment1}
    &\overline{T}= \frac{c  \ f^4(0)}{48 \pi f''(0)}\Big(  1-e^{- v^2 f''(0)t} \Big)\\
     &\overline{T^2} =\frac{c^2}{3456 \pi^2} \Bigg[ \frac{1}{2} \Big(\frac{f^{(4)}(0)}{f''(0)} \Big)^2(e^{-6 v^2 f''(0) t}- 6e^{- v^2 f''(0)t}+5)
     +\frac{f^{(6)}(0)}{f''(0)}(e^{-6 v^2 f''(0)t}-1) \Bigg]
\end{align}

\section{Free fermion dynamics}
\subsection{Evolution equation for the Wigner function}

Here, we will derive the evolution equation Eq.~\eqref{eq:ndephhomo1} for the Wigner function. 
We consider a model of spinless non-interacting fermions in one dimension. Let us first consider the Hamiltonian in the absence of the noise
\begin{equation}
\label{eq:Hff0}
\hat{H}_0 = \sum_i \hat{h}_i \;. \qquad \hat{h}_i = -J (\hat{a}^\dag_{i+1} \hat{a}_{i} + \hat{a}^\dag_{i} \hat{a}_{i+1}  - \mu \hat{a}^\dag_i \hat{a}_i )
\end{equation}
where we have introduced a chemical potential $\mu$ (chosen to be zero in the text).

In general, it is useful to represent densities which are conserved under the $H_0$ evolution with support around the site $j$ in the following form
\begin{equation}
\label{eq:defzj} 
    \hat{z}_j = \sum_{j'} \; \zeta_{j'} \;\hat{a}^\dag_{j + j'} \hat{a}_{j} =
\int_{-\pi}^{\pi} \frac{dk}{2\pi} \sum_{j'} e^{- \imath k j'} \zeta(k) \hat{a}^\dag_{j + j'} \hat{a}_{j} \;, \quad \left[\sum_j \hat{z}_j, \hat{H}_0\right] = 0
\end{equation}
where $\zeta(k)= \sum_j \zeta_j e^{i k j}$ is the Fourier transform of $\zeta_j$ and consider the coupling
\begin{equation}
\label{eq:Hff1}
\hat{H} = \hat{H}_0 + \sum_j \eta_j(\tau) (\hat{z}_j + \hat{z}^\dag_j )
\end{equation}
where the correlation of the noise $\eta_j(\tau)$ is defined in the main text.
We will keep $\zeta(k)$ arbitrary and specify its value only at the end. 

One can introduce the Wigner function which completely characterizes all correlation functions in any gaussian state.  Here, we will focus on the  noise average Wigner function which is defined as
\begin{equation}
\label{eq:ndef1}
n_\tau(k) \equiv \sum_{j'} e^{\imath k j'} \overline{\langle \hat{a}^{\dag}_{j + j'}  \hat{a}_{j}  \rangle_{\tau} } =  \frac 1 L \sum_{jj'} e^{\imath k j'} \Tr[ \hat{a}^{\dag}_{j + j'} \hat{a}_{j} \overline{\varrho_\tau}] \;.
\end{equation}
where $\langle \ldots \rangle_{\tau} = \braket{\Psi_0(\tau)  | \ldots | \Psi_0(\tau)}$ denotes the quantum average at time $\tau$ and we introduced the density matrix $\varrho_\tau \equiv \ket{\Psi_0(\tau)}\bra{\Psi_0(\tau)}$. Although we are interested in the groundstate of $\hat{H}_0$, these considerations apply to any translational invariant gaussian initial state $\ket{\Psi_0}$. In this case, because of the noise average, Eq.~\eqref{eq:ndef1} is independent of the position $j$.

We now consider the quantum evolution of $\varrho_\tau$.  It is easy to verify that, after the noise average, one obtains the Lindblad form
\begin{equation}
\label{SSE}
\frac{d \overline{\varrho_\tau}}{d\tau} = 
- \imath [H_0, \overline{\varrho_\tau}]  - \frac{\tau_0}{2} \sum_{j,j'} F(j-j') [\hat{z}_j + \hat{z}_j^\dag, [\hat{z}_{j'    } + \hat{z}_{j'}^\dag, \overline{\varrho_\tau}]]  
\end{equation}

We can use \eqref{eq:ndef1} and Eq.~\eqref{SSE}  to obtain an evolution equation for $n_\tau(k)$. Note that 
the first term in \eqref{SSE} does not contribute because of translational invariance and \eqref{eq:defzj}. 
Using \eqref{eq:defzj} we obtain explicitly 
\begin{multline}
\partial_\tau n_\tau(k)= 
- \frac{\tau_0}{2} \sum_{n,n', m, j,j'} F(n-n') e^{i k m} \left(\zeta_j \zeta_{j'} \overline{ 
\left\langle  [a_{n + j}^\dag \hat{a}_{n}, [a_{n' + j'}^\dag \hat{a}_{n'}, a^{\dag}_{m} \hat{a}_0]]\right\rangle_\tau }+\right. \\
\left.
+ \zeta_j^\ast \zeta_{j'}^\ast \overline{ 
\left\langle  [a_{n}^\dag \hat{a}_{n+j}, [a_{n'}^\dag \hat{a}_{n' + j'}, a^{\dag}_{m} \hat{a}_0]]\right\rangle_\tau } 
+ \zeta_j^\ast \zeta_{j'} \overline{ 
\left\langle  [a_{n}^\dag \hat{a}_{n+j}, [a_{n'+j'}^\dag \hat{a}_{n'}, a^{\dag}_{m} \hat{a}_0]]\right\rangle_\tau} + 
\zeta_j \zeta_{j'}^\ast \overline{ 
\left\langle  [a_{n+j}^\dag \hat{a}_{n}, [a_{n'}^\dag \hat{a}_{n'+j'}, a^{\dag}_{m} \hat{a}_0]]\right\rangle_\tau}
\right)
\end{multline}
We can expand all the commutators applying twice
\begin{equation}
    [ \hat{a}^\dag_{j_1} \hat{a}_{j_2}, a^{\dag}_{j_3} \hat{a}_{j_4}] = \delta_{j_2 j_3} \hat{a}^\dag_{j_1} \hat{a}_{j_4} - \delta_{j_1 j_4}\hat{a}^\dag_{j_3} \hat{a}_{j_2} \;.
\end{equation}
from the anti-commutation rules $\{ \hat{a}^\dag_{i} , \hat{a}^\dag_{j}\}=0$, $\{ \hat{a}^\dag_{i} , \hat{a}_{j}\}=\delta_{ij}$, and one gets
\be 
[ \hat{a}^\dag_{j_1} \hat{a}_{j_2}, [ a^{\dag}_{j_3} \hat{a}_{j_4}, a^{\dag}_{j_5} \hat{a}_{j_6}] =
\delta_{j_2 j_3} \delta_{j_4 j_5} a^{\dag}_{j_1} \hat{a}_{j_6}- 
\delta_{j_1 j_6} \delta_{j_4 j_5} a^{\dag}_{j_3} \hat{a}_{j_2}
- \delta_{j_3 j_6} \delta_{j_2 j_5} a^{\dag}_{j_1} \hat{a}_{j_4}
+ \delta_{j_3 j_6} \delta_{j_1 j_4} a^{\dag}_{j_5} \hat{a}_{j_2}
\ee 
As a result, a single correlator survives.
This is expected since the full Hamiltonian \eqref{eq:Hff1} is quadratic, so the correlation matrix $\langle \hat{a}^\dag_j \hat{a}_{j'} \rangle $ must satisfy a closed and linear evolution equation. Next one obtains, using the parity of the noise correlation $F(x) = F(-x)$ and translational invariance
\bea 
&& \partial_\tau n_\tau(k) = - \frac{\tau_0}{2}
 \sum_{m,j,j'} e^{i k m} \bigg(2 ( \zeta_j \zeta_{j'} + \zeta^*_{-j} \zeta^*_{-j'} ) \\
 && \times (F(j)  - F(m+j) ) 
+ ( \zeta_j \zeta^*_{-j'} + \zeta^*_{-j} \zeta_{j'} ) (F(0) + F(j+j') - F(m) - F(m+j+j'))  \bigg)
\overline{\langle \hat{a}^{\dag}_{m+j + j'}  \hat{a}_{0}  \rangle_{\tau} } \nonumber 
\eea 
Using that $\zeta(k)^* = \sum_j e^{i k j} \zeta^*_{-j}$ we finally obtain 
\bea 
\label{eq:ndephhomogeneral}
&& \partial_\tau n_\tau(k) =   \tau_0  \int_q \tilde F(q) | \zeta(-k) + \zeta(-q-k)^* |^2 (n(k+q) - n(k) ) 
\eea 
where 
\be \label{eq:Ftildedef}
\tilde F(q) = \sum_j e^{i q j} F(j) 
\ee 

Consider first half-filling, i.e. $\mu=0$, as in the main text. To recover the coupling with the noise defined there, we must choose $\zeta(k) = -J e^{i k}$. With this choice,
\begin{equation}
    | \zeta(-k) + \zeta(-q-k)^* |^2 = \epsilon(k + q/2)^2
\end{equation}
and one recovers \eqref{eq:ndephhomo1} given in the main text. 

For finite $-2<\mu<2$, the Hamiltonian in \eqref{eq:Hff0} remains critical and described by a $c=1$ CFT. In order to preserve parity and time-reversal symmetry, a simple choice is $\zeta(k) = \epsilon(k)/2$ with $\epsilon(k) = - J (2 \cos k - \mu)$, which corresponds to the coupling to the noise of the form
\begin{equation}
\label{eq:Hff1}
\hat{H} = \hat{H}_0 + \frac{1}{2}\sum_j \eta_j(\tau) (\hat{h}_j + \hat{h}_{j-1})
\end{equation}
In this case, we obtain after replacing $\zeta(k) \to \epsilon(k)/2$ in \eqref{eq:ndephhomogeneral}
\bea 
\label{eq:ndephhomo2}
&& \partial_\tau n_\tau(k) =   \frac{\tau_0}{4}  \int_q \tilde F(q) (\epsilon(k) + \epsilon(q+k))^2 (n(k+q) - n(k) )  \;. 
\eea

\subsection{Scaling limit}

Here we study the evolution equation for the Wigner function \eqref{eq:ndephhomo2}. In the scaling limit of large $\xi$ described in the main text, we look for a solution which has the following scaling form around the two Fermi points
\begin{equation} \label{eq:scalingn} 
 n_\tau(k) \simeq \nn(\xi (k_f - k), \tau/\xi) + \nn(\xi (k + k_f), \tau/\xi)
\end{equation}
The initial condition which corresponds to the ground state reads
\begin{equation}
\label{eq:inicondn}
    n_{\tau=0}(k) = \Theta(k + k_f) - \Theta(k - k_f) = -1 + \Theta(k + k_f) + \Theta(k_f - k)
\end{equation}
which gives for the initial scaling function $\nn(p, t = 0) = -1/2 + \Theta(p)$. We can now derive directly an evolution equation for the scaling function $\nn(p,t)$. 
In order to do this we replace in Eq.~\eqref{eq:ndephhomo2} $\hat\zeta(k)$ with the energy dispersion relation $\epsilon(k)$,
which gives Eq.~\eqref{eq:ndephhomo1} in the text. Next, we inject the scaling form Eq.~\eqref{eq:scalingn} around each Fermi point (choosing $+k_f$) 
%
and replace $k = k_f - p/\xi, k' = - p'/\xi$. From 
\eqref{eq:noisecorrFF} and \eqref{eq:Ftildedef}, we obtain that at large $\xi$, $\tilde{F}(k) \simeq \xi \hat f(\xi k)$, which leads to
(using $\tau_0=\xi$) 
\begin{multline}
 \label{eq:ndephscal}
 \partial_t \nn(p;t) \simeq
    \frac{\xi^2}{4} \int_{-\pi \xi}^{\pi \xi}  \frac{dp'}{2\pi}
 \tilde{f}(p') 
    \left(\epsilon\bigl(k_f - \frac{p}{\xi}\bigr) + \epsilon\bigl(k_f - \frac{p + p'}{\xi}\bigr) \right)^2 
    %
    %
    (\nn(p+p'; t) -  \nn(p; t)) \stackrel{\xi \to \infty}{\longrightarrow} 
    \\ 
    v^2 \int_{-\infty}^{\infty}  \frac{dp'}{2\pi}
 \tilde{f}(p') 
     \left(p + \frac{p'}{2}\right)^2 (\nn(p+p'; t) -  \nn(p; t)) =  
     v^2 \int_{-\infty}^{\infty}  \frac{dq}{2\pi}
 \tilde{f}(p-q) 
     \left(\frac{q+p}{2}\right)^2 
     (\nn(q; t) -  \nn(p; t))
 \end{multline}

We can now compute the mean energy density with respect to the ground state $\eee_F(\tau)$, which is expanded as
\begin{align}
\label{eq:enelim}
  &{\eee}_F(\tau = t \xi) = \int_{-\pi}^\pi \frac{dk}{2\pi} \epsilon(k) [n_\tau(k) - n_0(k)] \simeq \frac{1}{\xi^2}(\tilde{\eee}^+(t) +\tilde{\eee}^-(t))  \\ 
  &\tilde{\eee}^+(t)  \equiv \lim_{\xi \to \infty} \xi \int_{-\infty}^\infty \frac{dp}{2\pi} \epsilon(k_f - \frac p \xi) [\nn(p, t) - \nn(p, 0)] = 
  - v\int_{-\infty}^{\infty} \frac{dp}{2\pi} p \; [\nn(p, t)- \nn(p, 0)]
 \end{align}
and similarly for $\eee_F^{-}(t)$ with $k\sim -k_f$.

 We now prove the validity of Eq.~\eqref{eq:FFenelimit} in the main text. We show that $\tilde{\eee}^+(t)$ satisfies a a first order differential equation.
Indeed, differentiating \eqref{eq:enelim} w.r.t. the variable $t$ and using \eqref{eq:ndephscal}, we obtain 
 \begin{equation}
 \label{eq:eplustder}
      \partial_t \tilde{\eee}^+(t) = 
  - v^3 \int_{-\infty}^{\infty} \frac{dp}{2\pi}   \frac{dq}{2\pi} \; p
 \tilde{f}(p-q) 
     \left(\frac{q+p}{2}\right)^2 (\nn(q; t) -  \nn(p; t)) = -\frac{v^3}{2} \int_{-\infty}^\infty \frac{dq}{2\pi} \frac{dp}{2\pi}
 \tilde{f}(p-q) (p-q) \left(\frac{q + p}{2} \right)^2 [\nn(q; t) - \nn(p; t)] 
 \end{equation}
where in the last equality we symmetrized the integrand with respect to the exchange $p \leftrightarrow q$. 
The integral in \eqref{eq:eplustder} is finite but to proceed further, we need to split the integral in two terms involving $\nn(q; t)$ and $\nn(p; t)$, which are both divergent. To avoid this issue, we change variables $p = q + u$ and integrate by parts with respect to $q$ using that $\left.\nn(q; t) - \nn(q + u; t)\right|_{q=-\infty}^{\infty} = 0$ to arrive at
\begin{equation}
      \partial_t \tilde{\eee}^+(t) = v^3 \int \frac{dq}{2\pi} \frac{du}{2\pi}
 \tilde{f}(u) u \frac{(q + u/2)^3}{6} (\nn'(q; t) - \nn'(q+u;t))
\end{equation}
where $\nn'(q; t) = \partial_q \nn (q, t)$. We can now split the integral into two finite terms and replace $q \to q -u$ in the second integral. This leads to
\begin{multline}
\label{eq:diffeeplus}
      \partial_t \tilde{\eee}^+(t) = v^3 \int \frac{dq}{2\pi} \frac{du}{2\pi}
 \tilde{f}(u) u \frac{1}{6}\left[(q + u/2)^3-(q - u/2)^3\right]\nn'(q; t) 
 =\\=  
 v^3\int \frac{du}{2\pi} \tilde{f}(u) \frac{u^4}{48\pi} \int dq \; \nn'(q; t) + \frac{ v^3}{2} \int \frac{dq}{2\pi} q^2 \nn'(q; t) \int \frac{du}{2\pi} \tilde{f}(u) u^2 = \frac{v^3 f^{(4)}(0)}{48 \pi} - v^2 f''(0) \tilde{\eee}^+(\tau)    
\end{multline}
where we used that
\begin{align}
&\int dq \; \nn'(q; t) = 1 \;, \qquad \int \frac{dq}{2\pi} \; q^2 \nn'(q; t) =\int \frac{dq}{2\pi} \; q^2 [\nn'(q; t)-\nn'(q; 0)] =   \frac{2}{v}  \tilde{\eee}^+(t) \label{eq:partintn}
\\
& \int \frac{du}{2\pi} \tilde{f}(u) u^2 = -f''(0) \;, \qquad
\int \frac{du}{2\pi} \tilde{f}(u) u^4 = f^{(4)}(0) \;.
\end{align}
Note that in \eqref{eq:partintn} we used the initial condition \eqref{eq:inicondn} which shows that $\nn'(q, 0) = \delta(q)$.
We can thus solve \eqref{eq:diffeeplus} with the initial condition $\tilde{\eee}^+(t = 0) = 0$ and obtain
\begin{equation}
\label{eq:eeetilde}
    \tilde{\eee}^+(t) = 
    \frac{f^{(4)}(0) v \left(1-e^{-v^2 f''(0) t}\right)}{48 \pi  f''(0)}
\end{equation}
Summing also the equivalent contribution from $\tilde{\eee}^-(t)$ and using the definition of $\theta = -v^2 f''(0)/2$, we recover \eqref{T1moment} setting the central charge $c=1$, 
which is expected from universality. Note that the present calculation is from first principles. In principle the full scaling function $\nn(p; t)$ should be universal, i.e. 
independent of the microscopic details of the discrete model, but depending on $f(x)$, but we have not attempted to obtain it analytically.

\vspace{1cm}

To better characterize the time evolution of the scaling function $\nn(p; t)$, it is also possible to look at the time evolution of the \textit{moments} defined by
\begin{equation}
    M_n(t) \equiv \int_{-\infty}^\infty \frac{dp}{2 \pi} \; p^n \nn'(p,t) = - n \int_{-\infty}^\infty \frac{dp}{2 \pi} \; p^{n-1} ( \nn(p,t) - \nn(p,0))   \;.
\end{equation}
and clearly one has from \eqref{eq:partintn} $M_0(t) = 1$ and $M_2(t) = 2 \tilde{\eee}^+(t)/v$.
For higher $n$, one can show that these moments satisfy a hierarchy of differential equations which connects each moment with the previous ones, having the same parity, i.e.
\begin{equation}
\partial_t M_n(t) = \sum_{k} \alpha_k M_{n - 2 k}(t) \;.
\end{equation}
Because of the conservation of the density, one has $M_1(t) = 0$ and similarly all odd moments $M_{2k + 1}(t)$ vanish.

One can thus interpret $M_2(t)$ as the width of the distribution $\nn'(p; t)$ and this provides an estimation for the breakdown of the CFT description at the level of average quantities. Indeed, the Fermi points of $k \sim \pm k_F$ independently broaden with time, up to a time $\tau^\ast(\xi)$, where their width $\sim \xi \sqrt{M_2}$ is comparable with their initial distance $\sim 2 k_F$. Using $M_2(t) = 2 \tilde{\eee}^+(t)/v$ and \eqref{eq:eeetilde} This leads to 
\be 
\tilde e^+(\tau^*/\xi) = v k_F^2 \xi^2  \quad \Rightarrow \quad \tau^\ast \sim \frac{\xi}{v^2 |f''(0)|} \ln \left(
\frac{48 \pi |f''(0)| k_f^2 \xi^2}{f^{(4)}(0)}
\right)
\ee


\end{document}